\documentclass[twocolumn,journal]{IEEEtran}
\PassOptionsToPackage{ruled}{algorithm2e}
\usepackage[T1]{fontenc}
\usepackage[latin9]{inputenc}
\usepackage{color}
\usepackage{array}
\usepackage{multirow}
\usepackage{algorithm2e}
\usepackage{amsmath}
\usepackage{amsthm}
\usepackage{amssymb}
\usepackage{graphicx}
\usepackage{rotating}

\makeatletter

\providecommand{\tabularnewline}{\\}

\theoremstyle{plain}
\newtheorem{thm}{\protect\theoremname}
\theoremstyle{plain}
\newtheorem{lem}[thm]{\protect\lemmaname}

\DeclareMathOperator{\maximize}{maximize}
\DeclareMathOperator{\minimize}{minimize}
\DeclareMathOperator{\st}{subject~to}
\DeclareMathOperator{\diag}{diag}

\DeclareMathOperator{\tr}{Tr}
\DeclareMathOperator{\rank}{rank}
\DeclareMathOperator{\vect}{vec}

\newcommand{\herm}{^{{\dagger}}}
\newcommand{\trans}{^{\mathsf{T}}}
\DeclareMathOperator{\argmin}{argmin}

\usepackage{cite}
\usepackage{acronym}
\AtBeginDocument{\acrodef{AO}{alternating optimization}
\acrodef{APGM}{alternating projected gradient method}
\acrodef{APM}{accelerated proximal gradient method}
\acrodef{ASP}{antenna separation product}
\acrodef{AWGN}{additive white Gaussian noise}
\acrodef{BC}{broadcast channel}
\acrodef{BCM}{block coordinate maximization}
\acrodef{BEP}{bit error probability}
\acrodef{BER}{bit error rate}
\acrodef{BF-MIMO}[BF\mbox{-}MIMO]{beamforming MIMO}
\acrodef{BF}{beamforming}
\acrodef{BS}{base station}
\acrodef{bpcu}{bits per channel use}
\acrodef{CP}{cyclic prefix}
\acrodef{CSI}{channel state information}
\acrodef{CSIR}{channel state information at RX}
\acrodef{SSK}{space shift keying}
\acrodef{CSIT}{channel state information at TX}
\acrodef{DCMC}{discrete\mbox{-}input continuous\mbox{-}output memoryless channel}
\acrodef{DFT}{discrete Fourier transform}
\acrodef{DL-TR-GSM}{dual-layered transmit-receive \acl{GSM}}
\acrodef{DLT}{dual-layered transmission}
\acrodef{DPC}{dirty paper coding}
\acrodef{EGC}{equal gain combining}
\acrodef{EM}{electromagnetic}
\acrodef {EVD}{eigenvalue decomposition}
\acrodef{FSPL}{free space path loss}
\acrodef{FFT}{fast Fourier transform}
\acrodef{FDE}{frequency domain equalization}
\acrodef{GRSM}{generalized \acl{RSM}}
\acrodef{GSM}{generalized \acl{SM}}
\acrodef{IFFT}{invserse fast Fourier transform}
\acrodef{ICI}{inter-channel interference}
\acrodef{iid}[i.i.d.]{independent and identically distributed}
\acrodef{IQ}{in\mbox{-}phase and quadrature}
\acrodef{ISI}{intersymbol interference}
\acrodef{ISI-free}[ISI\mbox{-}free]{intersymbol interference free}
\acrodef{LIS}{large intelligent surface}
\acrodef{LOS}{line\mbox{-}of\mbox{-}sight}
\acrodef{KKT}{Karush\mbox{-}Kuhn\mbox{-}Tucker} 
\acrodef{MAC}{multiple-access channel}
\acrodef{mmWave}{millimeter-wave}
\acrodef{MIMO}{multiple\mbox{-}input multiple\mbox{-}output}
\acrodef{MISO}{multiple\mbox{-}input single\mbox{-}output}
\acrodef{ML}{maximum likelihood}
\acrodef{MRC}{maximal ratio combining}
\acrodef{MMSE}{minimum mean square error}
\acrodef{MU-TR-GSM}{multiuser transmit-receive  \acl{GSM} }
\acrodef{NCSIT}{no channel state information at TX}
\acrodef{NLOS}{non\mbox{-}\acs{LOS}} 
\acrodef{NOMA}{non-orthogonal multiple access}
\acrodef{OFDM}{orthogonal frequency division multiplexing}
\acrodef{OFDMA}{orthogonal frequency division multiple access}
\acrodef{PA}{power amplifier}
\acrodef{PAE}{power added efficiency}
\acrodef{PAPR}{peak\mbox{-}to\mbox{-}average power ratio}
\acrodef{PDF}{probability density function}
\acrodef{PEP}{pairwise error probability}
\acrodefplural{PEP}{pairwise error probabilities}
\acrodef{PGM}{projected gradient method}
\acrodef{PMP}{probability mass function}
\acrodef{PSM}{precoding-aided spatial modulation}
\acrodef{QSM}{quadrature spatial modulation}
\acrodef{RC}{reorganization computation}
\acrodef{RIS}{reconfigurable intelligent surface}
\acrodef{RSM}{receive spatial modulation}
\acrodef{RX}{receiver}
\acrodef{SEP}{symbol error probability}
\acrodef{SER}{symbol error rate}
\acrodef{SIC}{successive interference cancellation}
\acrodef{SINR}{signal-to-interference-plus-noise ratio}
\acrodef{SISO}{single-input single-output}
\acrodef{SM}{spatial modulation}
\acrodef{SMX-MIMO}[SMX\mbox{-}MIMO]{spatial multiplexing MIMO}
\acrodef{SMX}{spatial multiplexing}
\acrodef{SNR}{signal-to-noise ratio}
\acrodef{SC}{single carrier}
\acrodef{SVD}{singular value decomposition}
\acrodef{SPST}{single pole single-throw}
\acrodef{SU}{secondary user}
\acrodef{TDE}{time domain equalization}
\acrodef{TX}{transmitter}
\acrodef{ULA}{uniform linear array}
\acrodef{URA}{uniform rectangular array}
\acrodef{VGA}{variable gain amplifier}
\acrodef{ZF}{zero-forcing}
\acrodef{ZMCG}{zero-mean complex Gaussian}

}

\usepackage{pgfplots}
\usepackage{grffile}
\pgfplotsset{compat=newest}
\usetikzlibrary{plotmarks}
\usepgfplotslibrary{patchplots}
\usepackage{amsmath}

\definecolor{mycolor1}{rgb}{1.00000,0.00000,1.00000}%

\pgfplotsset{compat=newest}
\usetikzlibrary{plotmarks}
\usepgfplotslibrary{patchplots}

\tikzset{every node/.style={font=\small}}
\tikzset{every pin/.style={fill=white,font=\small}}
\tikzset{every pin edge/.style={<-,>=stealth,black,thick}}
\pgfplotsset{grid style={dotted,gray}}
\pgfplotsset{every axis/.style={inner sep=2pt}}
\pgfplotsset{legend style={font=\small}}
\newlength\figurewidth
\newlength\figureheight

\definecolor{mycolor1}{rgb}{1.00000,0.00000,1.00000}%
\newcommand{\algref}[1]{\textbf{Algorithm~\ref{#1}}}
\DeclareMathOperator{\argmax}{argmax}

\@ifundefined{showcaptionsetup}{}{%
 \PassOptionsToPackage{caption=false}{subfig}}
\usepackage{subfig}
\makeatother

\providecommand{\lemmaname}{Lemma}
\providecommand{\theoremname}{Theorem}

\begin{document}
\title{On the Maximum Achievable Sum-rate of the RIS-aided MIMO Broadcast
Channel}
\author{Nemanja~Stefan~Perovi\'c, \IEEEmembership{Member, IEEE}, Le-Nam
Tran, \IEEEmembership{Senior Member, IEEE},\\ Marco~Di~Renzo,
\IEEEmembership{Fellow, IEEE}, and Mark~F.~Flanagan, \IEEEmembership{Senior Member, IEEE}\thanks{Parts of this paper were presented at the IEEE International Workshop
on Signal Processing Advances in Wireless Communications (SPAWC),
Lucca, Italy, September 2021 \cite{Stefan:2021:multiuserRIS}.}
\thanks{\textcolor{black}{The work of N. S. Perovi\'c was supported by the European Commission through the H2020 SURFER project under grant agreement number 101030536.
	The work of Mark F. Flanagan was supported by the Irish Research Council under Grant IRCLA/2017/209.
	The work of L. N. Tran was supported in part by a Grant from Science
	Foundation Ireland under Grant number 17/CDA/4786. 
	The work of M. Di Renzo was supported in part by the European Commission through the H2020 ARIADNE project under grant agreement number 871464 and through the H2020 RISE-6G project under grant agreement number 101017011.}}
\thanks{N.~S.~Perovi\'c and M. Di Renzo are with Universit\'e Paris-Saclay,
CNRS, CentraleSup\'elec, Laboratoire des Signaux et Syst\`emes,
3 Rue Joliot-Curie, 91192 Gif-sur-Yvette, France. Email: nemanja-stefan.perovic@centralesupelec.fr,
marco.di-renzo@universite-paris-saclay.fr.}\thanks{L.-N. Tran and M. F. Flanagan are with the School of Electrical and
Electronic Engineering, University College Dublin, Belfield, Dublin
4, D04~V1W8, Ireland. Email: nam.tran@ucd.ie, mark.flanagan@ieee.org.}}
\maketitle
\begin{abstract}
Reconfigurable intelligent surfaces (RISs)\acused{RIS} represent
a new technology that can shape the radio wave propagation and thus
offers a great variety of possible performance and implementation
gains. Motivated by this, we investigate the achievable sum-rate
optimization in a \ac{BC} in the presence of \acp{RIS}. We solve
this problem by exploiting the well-known duality between the Gaussian
\ac{MIMO} \ac{BC} and the \ac{MAC}, and we correspondingly derive
three algorithms which optimize the users' covariance matrices and
the RIS phase shifts in the dual MAC. The users' covariance matrices
are optimized by a dual decomposition method with \ac{BCM}, or by
a gradient-based method. The RIS phase shifts are either optimized
sequentially by using a closed-form expression, or are computed in
parallel by using a gradient-based method. We present a computational
complexity analysis for the proposed algorithms. Simulation results
show that the proposed algorithms tend to converge to the same achievable
sum-rate overall, but may produce different sum-rate performance for
some specific situations, due to the non-convexity of the considered
problem. Also, the gradient-based optimization methods are generally
more time efficient. In addition, we demonstrate that the proposed
algorithms can provide a significant gain in the RIS-assisted \ac{BC}
assisted by multiple \acp{RIS} and that the gain depends on the placement
of the \acp{RIS}.\acresetall{}
\end{abstract}


\section{Introduction}

\bstctlcite{BSTcontrol}The need to satisfy constantly
increasing data rate demands in wireless communication networks motivates
the development of new technology solutions such as \acp{RIS}. An
\ac{RIS} is a metasurface that consists of a large number of small,
low-cost, and passive elements, as well as low-power electronic circuits
such as diodes or varactors. Since each of these elements can reflect
the incident signal with an adjustable phase shift, an RIS can effectively
shape the propagation of the impinging waves \cite{di2019smart,di2020smart}.
Therefore, the introduction of \acp{RIS} offers a wide variety of
possible implementation gains and potentially presents a new milestone
in wireless communications.

In order to fully exploit the gains that arise from the use of \acp{RIS},
we need to obtain a deep understanding of different aspects of RIS-assisted
wireless communication systems. Probably the most important aspect
is concerned with the optimal design of the RIS phase shifts, so that
the incoming radio wave is altered in a way that maximizes the aforementioned
gains. In this regard, the development of algorithms for optimizing
the achievable rate is of particular interest for \mbox{\ac{RIS}-aided}
communications. A significant body of research work in this area concentrates
on optimizing the achievable rate for point-to-point \ac{MIMO} communications.
The algorithms proposed in \cite{perovic2020achievable} and \cite{zhang2019capacity}
provide efficient methods for optimizing the transmit covariance matrix;
however, these works do not deal with multi-user \ac{MIMO}. The optimization
of the achievable rate for a single-stream \ac{MIMO} system in an
indoor \ac{mmWave} environment with a blocked direct link was analyzed
in \cite{perovic2019channel}. The optimization schemes proposed in
\cite{perovic2019channel} provide a near-optimal achievable rate
and require a low computational and hardware complexity. As far as
discrete signaling is concerned, the authors of \cite{perovic2020optimization}
have demonstrated that the achievable rate in RIS-aided systems can
be efficiently maximized by using the cutoff rate as a more tractable
optimization metric. The spectral efficiency enhancement arising from
the addition of a small number of active elements to the RIS was considered
in \cite{nguyen2021spectral}.

The optimization of the sum-rate in multi-user RIS-aided systems has
received increasing research attention as well. In \cite{wu2019intelligent},
the authors introduced an optimization method that increases the receive
\ac{SNR} and consequently enhances the achievable rate in \ac{MISO}
systems. The proposed solution is based on the \ac{AO} method, which
adjusts the transmit beamformer and the RIS phase shifts in an alternating
fashion. The \ac{AO} technique has also been successfully utilized
to increase the data rate for secure communications in environments
with multiple RISs and single-antenna users \cite{yu2020robust}.
In \cite{kammoun2020asymptotic}, the authors employed a gradient-based
algorithm to enhance the receive \ac{SINR}, and hence the achievable
rate, for single-antenna users that do not have a direct link with
the \ac{BS}. The sum-rate optimization for multi-user downlink communications
based on a deep reinforcement learning based algorithm was introduced
in \cite{huang2020reconfigurable}{.}{{}
In \cite{zhi2021two}, the authors derived an expression for the ergodic
achievable rate that depends on the statistical \ac{CSI}. As a result,
configuring the RIS in \cite{zhi2021two} only requires knowledge
of the \ac{CSI} statistics, which are assumed to change slowly. An
analytical framework for analyzing and optimizing the uplink and downlink
transmissions of \mbox{RIS-assisted} cell-free massive MIMO systems
when spatial correlation is present among the elements of the RIS
was introduced in \cite{van2021reconfigurable}. In \cite{ni2021resource},
the achievable sum-rate in a multi-cell \ac{NOMA} network was optimized
with respect to different network resources such as user association,
subchannel assignment, power allocation, phase shift design, and decoding
order.}

The aforementioned papers consider single-antenna user devices in
multi-user RIS-aided communications. On the other hand, a relatively
small number of papers study the use of multi-antenna user devices
in multi-user RIS-aided communications. This is due to the high intractability
of the resulting optimization problems. The use of an RIS in multi-cell
\ac{MIMO} systems was investigated in \cite{pan2020multicell} with
the aim of improving the weighted sum-rate, in particular for application
to the downlink transmission of cell-edge users. Because of the inherent
non-convexity of the optimization problem, it was first reformulated
and then solved by using the block coordinate descent (BCD) algorithm,
according to which the precoding matrices and the RIS phase shifts
are alternately optimized. Replacing some \acp{BS} with \acp{RIS}
in a multi-user MIMO cell-free network with multi-carrier transmission
was studied in \cite{zhang2021joint}. For the considered system,
the authors proposed an \ac{AO}-based optimization method, which
takes the specific features of multi-carrier transmission into account
for maximizing the weighted sum-rate. The achievable rate optimization
in cell-free networks with multiple \acp{BS} and \acp{RIS} was studied
in \cite{he2020multiple}. Therein, optimization algorithms for the
\ac{BS} transmit beamforming matrices and the RIS phase shifts were
separately derived, and later combined in an alternating {manner.
An \ac{AO}-based algorithm for maximizing a closed-form expression
for the asymptotic ergodic sum-rate in an RIS-aided \ac{MIMO} \ac{MAC}
without a direct link between users and the \ac{BS} was presented
in \cite{xu2021sum}. An \ac{AO} algorithm for maximizing the global
energy efficiency for uplink transmission when only partial \ac{CSI}
is known was proposed in \cite{you2021reconfigurable}. More precisely,
statistical \ac{CSI} was used for resource allocation in the considered
multi-user MIMO uplink networks, under the assumption that all the
signals are transmitted to the BS only via the RIS. The achievable
sum-rate optimization based on {\emph{a priori}}{{}
statistical knowledge of the users' locations for computing the phase
shifts of the \ac{RIS} elements was introduced in \cite{abrardo2021intelligent}.}{{}
}In \cite{ning2021terahertz}, the authors introduced an algorithm
for optimizing the \acp{RIS} and the hybrid-structured precoders/combiners,
in addition to a corresponding channel estimation method, for application
to an RIS-aided network operating in the Terahertz frequency~band.

All of the aforementioned papers assume, however, linear transmit
beamforming/precoding, which does not necessarily achieve the capacity
of the \ac{BC}. On the other hand, \ac{DPC} is an efficient technique
for achieving the channel capacity in the MIMO \ac{BC}. In \cite{Caire:ZFDPC:2003},
it was shown that implementing \ac{DPC} in a BC achieves the maximum
sum-rate. However, the analysis in \cite{Caire:ZFDPC:2003} was constrained
to a broadcast communication system with only two single-antenna user
terminals. The work in \cite{Caire:ZFDPC:2003} was extended to the
case with multiple users equipped with multiple antennas in \cite{yu2001trellis}.
A duality between the capacity region of a MIMO system with \ac{DPC}
in a BC and the capacity region of the MIMO \ac{MAC} was established
in \cite{Vishwanath:duality_achievable:2003}. Accordingly, the capacity
region of a MIMO \ac{BC} with \ac{DPC} was proved to be the same
as the capacity region of the dual MIMO \ac{MAC}, under the assumption
that the transmitters have the same sum power constraint as the MIMO
\ac{BC}. Utilizing this duality, the authors proposed simple and
fast iterative algorithms that provide the sum capacity achieving
strategy for the dual MAC, which can easily be converted to the equivalent
optimal strategies for the BC \cite{Jindal:IterativeWF:BC:2005}.
An application of the \ac{BC}-\ac{MAC} duality to a multi-user \ac{MISO}
system was studied in \cite{Nam:beamdesign:ZFDPC:2012}. More precisely,
the duality between a BC with \ac{ZF}-DPC and a MAC with ZF-based
\ac{SIC} was used to design the transmit beamformer. To the best
of the authors' knowledge, the only paper that exploits the \ac{BC}-\ac{MAC}
duality for studying the capacity/achievable rate regions for the
\ac{MAC} and for the \ac{BC} in RIS-aided communications is \cite{zhang2020intelligent}.
However, the analysis presented in \cite{zhang2020intelligent} was
limited to single-antenna user terminals and a single-antenna \ac{BS},
and can not be directly extended to multi-antenna devices.

The contributions of this paper are listed as follows:
\begin{itemize}
\item We exploit the Gaussian MIMO \ac{BC}-\ac{MAC} duality to~maximize
the achievable sum-rate of a multi-user MIMO system equipped with
\acp{RIS} communicating over a BC, and formulate an optimization
problem of the users'~covariance matrices and the phase shifts of
the RIS~elements.
\item Due to the non-convexity of the optimization problem and the possibility
that a local optimization method may be trapped in a bad local optimum,{{}
we propose three different iterative algorithms which operate in the
dual MAC, each of which provides a locally optimum solution. }The
first algorithm, which we call the \emph{AO algorithm}, optimizes
the users' covariance matrices and the phase shifts of the RIS elements
in an alternating manner. The users' covariance matrices are obtained
by a dual decomposition method with a \ac{BCM}, while the phase shifts
of the RIS elements are computed sequentially and are formulated in
a closed-form expression. As it can be desirable to increase the time
efficiency of the aforementioned sequential optimization, we introduce
the \emph{approximate \ac{AO} algorithm}, which uses a gradient-based
method for optimizing simultaneously the phase shifts of the RIS elements.
Finally, the \emph{\ac{APGM} algorithm} applies a gradient-based
method for optimizing the users' covariance matrices and the phase
shifts of the RIS elements.
\item For the proposed algorithms, we provide the computational complexity
in terms of the number of complex multiplications.
\item We show through simulations that the proposed algorithms provide the
same achievable sum-rate with a low number of iterations, when the
degree of freedom is high. The \ac{AO} algorithm requires the least
number of iterations, but has the longest execution time. This is
mainly attributed to the sequential optimization of the phase shifts
of the RIS elements. The gradient-based optimization of the users'
covariance matrices and the phase shifts of the RIS elements for the
APGM algorithm is, on the other hand, the most efficient in terms
of {execution time. On the other hand, when the degree
of freedom is low, the proposed algorithms may yield different sum-rate
performance. In such cases, all the proposed algorithms need to be
executed so that the best achievable sum-rate is obtained with high
probability.}
\item We show that the achievable sum-rate increases approximately logarithmically
with the number of transmit antennas and the number of users in the
BC. {Also, we demonstrate that \ac{DPC} always provides
a larger achievable sum-rate than linear precoding and that the gains
increase with the number of RIS elements. }Moreover, we show that
substantial achievable sum-rate gains can be obtained in the multi-RIS
case and that these gains depend on the placement of the \acp{RIS}.
\end{itemize}
$\quad$The rest of this paper is organized as follows. In Section~\ref{sec:System-Model},
we introduce the system model of the considered \mbox{RIS-aided}
MIMO BC. In Section \ref{sec:Problem-Formulation}, we formulate the
optimization problem to maximize the achievable sum-rate. In Section
\ref{sec:Proposed-Optimization-Methods}, we propose and derive three
optimization algorithms to solve the formulated optimization problem.
The analysis of the computational complexity of the proposed optimization
algorithms is presented in Section \ref{sec:Computational-Complexity}.
In Section \ref{sec:Simulation-Results}, we provide simulation results
that illustrate the achievable sum-rate of the proposed algorithms.
Finally, Section \ref{sec:Conclusion} concludes this paper.

\textit{Notation}: Bold lower and upper case letters represent vectors
and matrices, respectively. $\mathbb{C}^{m\times n}$ denotes the
space of $m\times n$ complex matrices. $\mathbf{H}\trans$ and $\mathbf{H}\herm$
denote the transpose and Hermitian transpose of $\mathbf{H}$, respectively;
$|\mathbf{H}|$ is the determinant of $\mathbf{H}$. $\tr(\mathbf{H})$
denotes the trace of $\mathbf{H}$ and $\rank(\mathbf{H})$ denotes
the rank of $\mathbf{H}$. $\lambda_{\max}(\mathbf{H})$ denotes the
largest singular value of $\mathbf{H}$. $\log_{2}(\cdot)$ is the
binary logarithm, $\ln(\cdot)$ is the natural logarithm and $(x)_{+}$
denotes $\max(0,x)$. $\mathbb{E}\bigl\{\cdot\bigr\}$ denotes the
expectation operator and $\left(\cdot\right)^{\ast}$ denotes the
complex conjugate. $\left\Vert \mathbf{H}\right\Vert $ denotes the
Frobenius norm of $\mathbf{H}$ which reduces to the Euclidean norm
if $\mathbf{H}$ is a vector. $\vect_{d}(\mathbf{H})$ is the vector
comprised of the diagonal elements of $\mathbf{H}$. $P_{\mathcal{\mathcal{C}}}(\mathbf{u})$
denotes the Euclidean projection of $\mathbf{u}$ onto the set $\mathcal{C}$,
i.e., $P_{\mathcal{\mathcal{C}}}(\mathbf{u})=\arg\min_{\mathbf{x}\in\mathcal{\mathcal{C}}} ||\mathbf{x}-\mathbf{u}||$.
The notation $\mathbf{A}\succeq(\succ)\mathbf{B}$ means that $\mathbf{A}-\mathbf{B}$
is positive semidefinite (definite). $\mathbf{I}$ represents an identity
matrix whose size depends from the context. $\Re(\mathbf{x})$ and
$\Im(\mathbf{x})$ denote the real and imaginary part of $\mathbf{x}$,
respectively. For a vector~$\mathbf{x}$, $\diag(\mathbf{x})$ denotes
a diagonal matrix with the elements of $\mathbf{x}$ on the diagonal.
$\mathcal{CN}(\mu,\sigma^{2}$) denotes a circularly symmetric complex
Gaussian random variable with mean $\mu$ and variance~$\sigma^{2}$.
The symbol $\odot$ denotes the Hadamard product, i.e., the element-wise
product, of two matrices. $|x|$ denotes the modulus of the complex
number $x$, and $|\mathbf{x}|$, $\mathbf{x}\in\mathbb{C}^{N\times1}$,
is defined as $|\mathbf{x}|=\begin{bmatrix}|x_{1}| & |x_{2}| & \cdots & |x_{N}|\end{bmatrix}\trans$.
Similarly, we define $\frac{1}{|\mathbf{x}|}=\begin{bmatrix}\frac{1}{|x_{1}|} & \frac{1}{|x_{2}|} & \cdots & \frac{1}{|x_{N}|}\end{bmatrix}\trans$.
Finally, we denote by $\nabla_{\mathbf{x}}f(\cdot)$ the complex gradient
of $f(\cdot)$ with respect to $\mathbf{x}^{\ast}$, i.e., $\nabla_{\mathbf{x}}f(\cdot)=\frac{1}{2}\Bigl(\frac{\partial f(\cdot)}{\partial\Re(\mathbf{x})}+j\frac{\partial f(\cdot)}{\partial\Im(\mathbf{x})}\Bigr)$.

\section{System Model\label{sec:System-Model}}

{We consider a \ac{BC} in which one \ac{BS} simultaneously
serves $K$ users, as shown in Fig. \ref{fig:Aerial-view}. Both the
\ac{BS} and the users are equipped with multiple antennas, such that
the \ac{BS} and the }{\emph{k}}{-th
user have $N_{t}$ and $n_{k}$ antennas, respectively. The \ac{BS}
antennas are placed in a \ac{ULA} with inter-antenna separation $s_{t}$.
In a similar manner, all the antennas of a single user are placed
in a \ac{ULA} with inter-antenna separation $s_{r}$. In order to
improve the system performance, $N_{s}$ \acp{RIS} are also present
in the considered communication environment. Each \ac{RIS} consists
of $N_{\mathrm{ris}}$ reflecting elements}\footnote{{To simplify the mathematical presentation, we assumed that
all RISs have the same number of reflecting elements, but the considered
system model and proposed algorithms are also applicable to case
where RISs have a different number of reflecting elements.}}{{} which are placed in a \ac{URA}}, so that the separation
between the centers of adjacent \ac{RIS} elements in both dimensions
is $s_{\mathrm{ris}}$.
\begin{figure}[t]
\includegraphics[width=8.84cm]{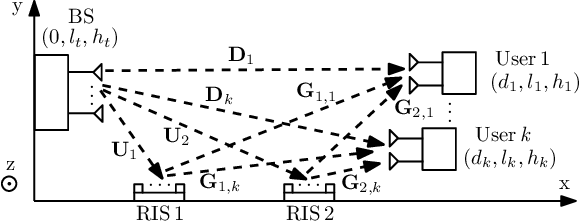}{\caption{{Aerial view of the considered communication system
for the case of 2 RISs.\label{fig:Aerial-view}}}
}
\end{figure}

The received signal at the \emph{k}-th user is given by 
\begin{align}
\mathbf{y}_{k} & =\mathbf{H}_{k}\mathbf{x}_{k}+\sum\limits _{j=1,j\neq k}^{K}\mathbf{H}_{k}\mathbf{x}_{j}+\mathbf{n}_{k}\label{eq:sigmod:gen}
\end{align}
where $\mathbf{H}_{k}\in\mathbb{C}^{n_{k}\times N_{t}}$ is the channel
matrix for the \emph{k}-th user, $\mathbf{x}_{k}\in\mathbb{C}^{N_{t}\times1}$
is the transmitted signal intended for the \emph{k}-th user, and $\mathbf{x}_{j}\in\mathbb{C}^{N_{t}\times1}$
for $j\neq k$ are the transmitted signals intended for the other users,
which act as interference for the detection of $\mathbf{x}_{k}$.
The noise vector $\mathbf{n}_{k}\in\mathbb{C}^{n_{k}\times1}$ consists
of \ac{iid} elements that are distributed according to $\mathcal{CN}(0,N_{0})$,
where $N_{0}$ is the noise variance. The channel matrix for the $k$-th
user can be written~as
\begin{equation}
\mathbf{H}_{k}=\mathbf{D}_{k}+\sum\limits _{i=1}^{N_{s}}\mathbf{G}_{i,k}\mathbf{F}(\boldsymbol{\theta}_{i})\mathbf{U}_{i}\label{eq:Hk_equ-1-1}
\end{equation}
where $\mathbf{D}_{k}\in\mathbb{C}^{n_{k}\times N_{t}}$ is the direct
link channel matrix between the \ac{BS} and the \emph{k}-th user,
$\mathbf{U}_{i}\in\mathbb{C}^{N_{\mathrm{ris}}\times N_{t}}$ is the
channel matrix between the BS and the $i$-th RIS, and $\mathbf{G}_{i,k}\in\mathbb{C}^{n_{k}\times N_{\mathrm{ris}}}$
is the channel matrix between the \emph{i}-th RIS and the \emph{k}-th
user. The signal reflection from the \emph{i}-th RIS is modeled by
$\mathbf{F}(\boldsymbol{\theta}_{i})=\mathrm{diag}(\boldsymbol{\theta}_{i})\in\mathbb{C}^{N_{\mathrm{ris}}\times N_{\mathrm{ris}}}$,
where $\boldsymbol{\theta}_{i}=[\theta_{i,1},\theta_{i,2},\ldots,\theta_{i,N_{\mathrm{ris}}}]\trans\in\mathbb{C}^{N_{\mathrm{ris}}\times1}$.
For mathematical convenience, we equivalently rewrite the channel
matrix $\mathbf{H}_{k}$ in a compact form as
\begin{equation}
\mathbf{H}_{k}=\mathbf{D}_{k}+\mathbf{G}_{k}\mathbf{F}(\boldsymbol{\theta})\mathbf{U}\label{eq:Hk_equ}
\end{equation}
where $\mathbf{G}_{k}=[\mathbf{G}_{1,k}\;\mathbf{G}_{2,k}\;\cdots\;\mathbf{G}_{N_{s},k}]\in\mathbb{C}^{n_{k}\times N_{s}N_{\mathrm{ris}}}$,
$\mathbf{U}=[\mathbf{\mathbf{U}}_{1}\trans\;\mathbf{\mathbf{U}}_{2}\trans\;\cdots\;\mathbf{\mathbf{U}}_{N_{s}}\trans]\trans\in\mathbb{C}^{N_{s}N_{\mathrm{ris}}\times N_{t}}$,
$\boldsymbol{\theta}=[\boldsymbol{\theta}_{1}\trans\;\boldsymbol{\theta}_{2}\trans\;\ldots\;\boldsymbol{\theta}_{N_{s}}\trans]\trans\in\mathbb{C}^{N_{s}N_{\mathrm{ris}}\times1}$
and $\mathbf{F}(\boldsymbol{\theta})=\diag(\mathbf{\boldsymbol{\theta}})$.
We assume that the signal reflection from any RIS element is ideal
(i.e., without any power loss) and therefore we may write $\theta_{l}=e^{j\phi_{l}}$
for $l=1,2,\ldots,N_{s}N_{\mathrm{ris}}$, where $\phi_{l}$ is the
phase shift induced by the $l$-th RIS element. Equivalently, this
can be written as
\begin{equation}
|\boldsymbol{\theta}|=1\Leftrightarrow\left|\theta_{l}\right|=1,\quad l=1,2,\ldots,N_{s}N_{\mathrm{ris}}.\label{eq:RIS_elem_cons}
\end{equation}

\section{Problem Formulation\label{sec:Problem-Formulation}}

In this paper, we are interested in maximizing the achievable sum-rate
of the considered RIS-assisted wireless communication system. To accomplish
this, we exploit the fact that the achievable rate region of a Gaussian
MIMO BC can be achieved by \ac{DPC} \cite{Weingarten:CapacityRegion:MU_MIMO:2006}.
\ac{DPC} enables us to perfectly eliminate the interference term
$\sum_{j<k}\mathbf{H}_{k}\mathbf{x}_{j}$ for the \emph{k}-th user,
assuming that the BS has full (non-causal) knowledge of this interference
term. Let $\pi$ be an ordering of users, i.e., a permutation of the
set $\{1,2,\ldots,K\}$. Then for this ordering, the achievable rate
for the \emph{k}-th user can be computed as \cite[Eq. (3)]{Vishwanath:duality_achievable:2003}
\begin{equation}
R_{\pi(k)}=\log_{2}\frac{\Bigl|\mathbf{I}+\mathbf{H}_{\pi(k)}\bigl(\sum_{j\geq k}\mathbf{S}_{\pi(j)}\bigr)\mathbf{H}_{\pi(k)}\herm\Bigr|}{\Bigl|\mathbf{I}+\mathbf{H}_{\pi(k)}\bigl(\sum_{j>k}\mathbf{S}_{\pi(j)}\bigr)\mathbf{H}_{\pi(k)}\herm\Bigr|},k=1,\ldots,K
\end{equation}
where $\mathbf{S}_{k}=\mathbb{E}\bigl\{\mathbf{x}_{k}\mathbf{x}_{k}\herm\bigr\}\succeq\mathbf{0}$
is the input covariance matrix of user $k$. In this paper, we consider
a sum-power constraint at the BS, i.e.,
\begin{equation}
\sum\limits _{k=1}^{K}\tr\bigl(\mathbf{S}_{k}\bigr)\leq P
\end{equation}
where $P$ is the maximum total power at {the BS.
Therefore, the achievable rate optimization problem for the RIS-assisted
MIMO BC can be expressed as\begin{subequations}{\label{eq:MIMO:BS:sumrate}}
\begin{align}
\underset{\mathbf{S},\boldsymbol{\theta}}{\maximize} & \quad\sum\limits _{k=1}^{K}\log_{2}\frac{\Bigl|\mathbf{I}+\mathbf{H}_{\pi(k)}\bigl(\sum_{j\geq k}\mathbf{S}_{\pi(j)}\bigr)\mathbf{H}_{\pi(k)}\herm\Bigr|}{\Bigl|\mathbf{I}+\mathbf{H}_{\pi(k)}\bigl(\sum_{j>k}\mathbf{S}_{\pi(j)}\bigr)\mathbf{H}_{\pi(k)}\herm\Bigr|}\\
\st & \quad\sum\limits _{k=1}^{K}\tr\bigl(\mathbf{S}_{k}\bigr)\leq P;\mathbf{S}_{k}\succeq\mathbf{0},\forall k,\\
 & \quad|\boldsymbol{\theta}|=1,
\end{align}
{}\end{subequations}where $\mathbf{S}\triangleq(\mathbf{S}_{k})_{k=1}^{K}$.
It is w}orth mentioning that the achievable sum-rate in \eqref{eq:MIMO:BS:sumrate}
is independent of the ordering of users $\pi$ \cite{Vishwanath:duality_achievable:2003}.
We remark that the objective function of the above problem is neither
convex nor concave with the input covariance matrices and the phase
shifts, and thus directly solving \eqref{eq:MIMO:BS:sumrate} is difficult.
In \cite{Vishwanath:duality_achievable:2003}, Vishwanath \emph{et
al.} established what is now well-known as the \emph{BC-MAC duality},
and showed that the achievable sum-rate of the MIMO \ac{BC} equals
the achievable rate of the dual Gaussian MIMO \ac{MAC}. As a result,
\eqref{eq:MIMO:BS:sumrate} is \mbox{equivalent~to} \begin{subequations}\label{eq:MIMO:MAC:sumrate}
\begin{align}
\underset{\bar{\mathbf{S}},\boldsymbol{\theta}}{\maximize} & \quad f(\boldsymbol{\theta},\bar{\mathbf{S}})\triangleq\ln\Bigl|\mathbf{I}+\sum\limits _{k=1}^{K}\mathbf{H}_{k}\herm\bar{\mathbf{S}}_{k}\mathbf{H}_{k}\Bigr|\\
\st & \quad\bar{\mathbf{S}}\in\mathcal{S}\\
 & \quad\boldsymbol{\theta}\in\varTheta.
\end{align}
\end{subequations} where $\bar{\mathbf{S}}\triangleq(\bar{\mathbf{S}}_{k})_{k=1}^{K}$,
$\mathbf{H}_{k}\herm$ is referred to as the dual \ac{MAC} channel
corresponding to $\mathbf{H}_{k}$ and $\bar{\mathbf{S}}_{k}\in\mathbb{C}^{n_{k}\times n_{k}}$
is the input covariance matrix of user $k$ in the dual MAC. The sets
$\mathcal{S}$ and $\varTheta$ in \eqref{eq:MIMO:MAC:sumrate} are
defined as
\begin{align}
\mathcal{S} & =\{\bar{\mathbf{S}}\ |\ \sum\limits _{k=1}^{K}\tr\bigl(\bar{\mathbf{S}}_{k}\bigr)\leq P;\bar{\mathbf{S}}_{k}\succeq\mathbf{0}\thinspace\thinspace\forall k\}\\
\varTheta & =\bigl\{\boldsymbol{\theta}\in\mathbb{C}^{N_{s}N_{\mathrm{ris}}\times1}\ |\ \bigl|\boldsymbol{\theta}\bigr|=1\bigr\}.
\end{align}
Once the input covariance matrices $(\bar{\mathbf{S}}_{k})_{k=1}^{K}$
in the dual MAC are found, the corresponding covariance matrices $(\mathbf{S}_{k})_{k=1}^{K}$
in the BC are computed as \cite[Eq. (11)]{Vishwanath:duality_achievable:2003}
\begin{equation}
\mathbf{S}_{k}=\mathbf{B}_{k}^{-1/2}\mathbf{F}_{k}\mathbf{G}_{k}\herm\mathbf{A}_{k}^{1/2}\bar{\mathbf{S}}_{k}\mathbf{A}_{k}^{1/2}\mathbf{G}_{k}\mathbf{F}_{k}\herm\mathbf{B}_{k}^{-1/2}\label{eq:Cov_mat_rel}
\end{equation}
where $\mathbf{A}_{k}=\mathbf{I}+\mathbf{H}_{k}(\sum_{i=1}^{k-1}\mathbf{S}_{i})\mathbf{H}_{k}\herm$
and $\mathbf{B}_{k}=\mathbf{I}+\sum_{i=k+1}^{K}\mathbf{H}_{i}\herm\bar{\mathbf{S}}_{i}\mathbf{H}_{i}$,
and the \ac{SVD} of $\mathbf{B}_{k}^{-1/2}\mathbf{H}_{k}\herm\mathbf{A}_{k}^{-1/2}$
is $\mathbf{F}_{k}\boldsymbol{\Lambda}_{k}\mathbf{G}_{k}\herm$. We
also note that the expression for the MAC-BC conversion is obtained
under the assumption that the encoding ordering of the users in the BC
channel is from the last user to the first user. To make \eqref{eq:Cov_mat_rel}
applicable to the case of an arbitrary encoding ordering of users,
the index $k$ needs to be replaced with $\pi(k)$.

\section{Proposed Optimization Methods\label{sec:Proposed-Optimization-Methods}}

\subsection{Alternating Optimization (AO)}

To solve \eqref{eq:MIMO:MAC:sumrate}, we propose an efficient \ac{AO}
method, which adjusts the covariance matrices and the phase shifts
of the RIS elements in an alternating fashion. First, we propose an
iterative approach which optimizes all the covariance matrices
in the dual MAC by using a \ac{BCM} approach. Next, the optimal phase
shift for each RIS element is obtained using a closed-form
expression, similar to \cite{zhang2019capacity}.

\subsubsection{Covariance Matrix Optimization\label{subsec:AO-Cov-Mat}}

For a given $\boldsymbol{\theta}$, the achievable rate optimization
problem in \eqref{eq:MIMO:MAC:sumrate} is simplified as{\begin{subequations}{\label{eq:MIMO:MAC:fixtheta}}
\begin{align}
\underset{\bar{\mathbf{S}}}{\maximize} & \quad\ln\Bigl|\mathbf{I}+\sum\limits _{k=1}^{K}\mathbf{H}_{k}\herm\bar{\mathbf{S}}_{k}\mathbf{H}_{k}\Bigr|\\
\st & \quad\bar{\mathbf{S}}\in\mathcal{S}.\label{eq:MAC:SPC}
\end{align}
{}\end{subequations}}The above optimization problem
is convex and thus it can be solved by off-the-shelf convex solvers.
In this paper, we propose a more efficient method which combines the
dual decomposition method and accelerated block coordinate maximization
method to solve \eqref{eq:MIMO:MAC:fixtheta}. The details are given
next.

Following the dual decomposition method, we first form the partial
Lagrangian function of \eqref{eq:MIMO:MAC:fixtheta} as
\begin{equation}
\mathcal{L}(\mu,\bar{\mathbf{S}})=\ln\Bigl|\mathbf{I}+\sum_{k=1}^{K}\mathbf{H}_{k}\herm\bar{\mathbf{S}}_{k}\mathbf{H}_{k}\Bigr|-\mu\biggl[\sum_{k=1}^{K}\tr\bigl(\bar{\mathbf{S}}_{k}\bigr)-P\biggr]\label{eq:Lang_funct}
\end{equation}
where $\mu$ is the Lagrangian multiplier for the constraint in \eqref{eq:MAC:SPC}.
For mathematical convenience, we use the natural logarithm in \eqref{eq:Lang_funct}
without affecting the optimality of \eqref{eq:MIMO:MAC:fixtheta}.
For a given $\mu$, the dual function is given as
\begin{equation}
g(\mu)=\underset{\bar{\mathbf{S}}\succeq\mathbf{0}}{\max}\;\mathcal{L}\bigl(\mu,\bar{\mathbf{S}}\bigr)\label{eq:dual}
\end{equation}
where the constraint $\bar{\mathbf{S}}\succeq\mathbf{0}$ is understood
as $\bar{\mathbf{S}}_{k}\succeq\mathbf{0}$, $\forall k$. To evaluate
$g(\mu)$, in \cite{Stefan:2021:multiuserRIS} we have presented a
cyclic block maximization method which cyclically optimizes each $\bar{\mathbf{S}}_{k}$
while keeping the other $\bar{\mathbf{S}}_{j}$ $(j\ne k)$ fixed,
and which was first applied in \cite{Yu:SumCapacity:MIMO_BC:Decomposition:2006}
to a system without an RIS. For the purpose of exposition, let us
define $\bar{\mathbf{S}}^{(n)}\triangleq\bigl(\bar{\mathbf{S}}_{1}^{(n)},\ldots,\bar{\mathbf{S}}_{k-1}^{(n)},\bar{\mathbf{S}}_{k}^{(n)},\bar{\mathbf{S}}_{k+1}^{(n)},\ldots,\bar{\mathbf{S}}_{K}^{(n)}\bigr)$
which represents the current iterate. The $k$-th element $\bar{\mathbf{S}}_{k}$
of the next iterate is found to be the optimal solution of the following
problem:
\begin{align}
\!\!\!\underset{\bar{\mathbf{S}}_{k}\succeq\mathbf{0}}{\maximize} & \quad\ln\Bigl|\mathbf{I}+\bar{\mathbf{H}}_{k}^{-1/2}\mathbf{H}_{k}\herm\bar{\mathbf{S}}_{k}\mathbf{H}_{k}\bar{\mathbf{H}}_{k}^{-1/2}\Bigr|-\mu\tr\bigl(\bar{\mathbf{S}}_{k}\bigr)\label{eq:Sbarmax}
\end{align}
where
\begin{equation}
\bar{\mathbf{H}}_{k}=\mathbf{I}+\sum\limits\limits _{j=1,j\neq k}^{K}\mathbf{H}_{j}\herm\bar{\mathbf{S}}_{j}\mathbf{H}_{j}.\label{eq:Hbar_k}
\end{equation}
It can be seen that the optimal solution to \eqref{eq:Sbarmax} is
given by~\cite{Yu:SumCapacity:MIMO_BC:Decomposition:2006}
\begin{equation}
\bar{\mathbf{S}}_{k}^{\star}=\mathbf{V}_{k}\diag\Bigl(\Bigl[\bigl(\frac{1}{\mu}-\frac{1}{\sigma_{1}}\bigr)_{+},\bigl(\frac{1}{\mu}-\frac{1}{\sigma_{2}}\bigr)_{+},\ldots,\bigl(\frac{1}{\mu}-\frac{1}{\sigma_{r}}\bigr)_{+}\Bigr]\trans\Bigr)\mathbf{V}_{k}\herm\label{eq:Sk_opt}
\end{equation}
where $\mathbf{H}_{k}\bar{\mathbf{H}}_{k}^{-1}\mathbf{H}_{k}\herm=\mathbf{V}_{k}\diag\Bigl(\sigma_{1},\sigma_{2},\ldots,\sigma_{r}\Bigr)\mathbf{V}_{k}\herm$
is the \ac{EVD} of $\mathbf{H}_{k}\bar{\mathbf{H}}_{k}^{-1}\mathbf{H}_{k}\herm$
and $r=\rank(\mathbf{H}_{k})\le\min(N_{t},n_{k})$. The cyclic block
coordinate maximization (CBCM) method for solving \eqref{eq:dual}
is summarized in \algref{alg:CBCM}.
\begin{algorithm}[t]
{\small\caption{$\bar{\mathbf{S}}^{\star}\leftarrow\mathtt{CBCM}(\bar{\mathbf{S}}^{(0)})$\label{alg:CBCM}}

\SetKwInput{KwData}{Initialization}
\SetAlgoNoLine
\DontPrintSemicolon
\LinesNumbered

\KwIn{$\bar{\mathbf{S}}^{(0)}$}

Set $n\leftarrow0$\;

\Repeat{a stopping criterion is met}{

Compute $\bar{\mathbf{S}}_{k}^{(n+1)}$ according to \eqref{eq:Sk_opt}
for $k=1,2,\ldots,K$\;

$n\leftarrow n+1$\;

}

\KwOut{$\bar{\mathbf{S}}^{\star}=\bar{\mathbf{S}}^{(n)}$}}
\end{algorithm}

We remark that the solution to \eqref{eq:Sbarmax} is unique and thus
\algref{alg:CBCM} is guaranteed to converge to the optimal solution
of \eqref{eq:dual}. Since \algref{alg:CBCM} is then repeatedly used
to solve \eqref{eq:dual}, it is important to analyze its convergence
rate, which has not been studied previously. In this regard, the next
theorem is in order.
\begin{thm}
\label{thm:CBCM}Let $M=\max_{1\leq k\leq K}
\lambda_{\max}^{2}\bigl(\mathbf{H}_{k}\mathbf{H}_{k}\herm\bigr)$
and $R(\bar{\mathbf{S}}^{\mathrm{initial}})=\max_{\bar{\mathbf{S}}\succeq\mathbf{0}} \bigl\{||\bar{\mathbf{S}}-\bar{\mathbf{S}}^{\mathrm{initial}}||:\mathcal{L}\bigl(\mu,\bar{\mathbf{S}}\bigr)\geq\mathcal{L}\bigl(\mu,\bar{\mathbf{S}}^{\mathrm{initial}}\bigr)\bigr\}$.
Then we have 
\begin{equation}
\mathcal{L}_{\mu}^{\ast}-\mathcal{L}\bigl(\mu,\bar{\mathbf{S}}^{(n)}\bigr)\leq2cMK^{2}R^{2}\frac{1}{n},\;\forall n\geq1,
\end{equation}
where $c=\max\bigl(\frac{2}{MK^{2}R^{2}}-2,2,\mathcal{L}_{\mu}^{\ast}-\mathcal{L}(\mu,\bar{\mathbf{S}}^{(1)})\bigr)$
and $\mathcal{L}_{\mu}^{\ast}$ is the optimal objective of \eqref{eq:dual}.
\end{thm}
\begin{IEEEproof}
See Appendix \ref{subsec:Proof-of-TheoremCBCM}.
\end{IEEEproof}
Theorem \ref{thm:CBCM} indicates that the convergence rate of \algref{alg:CBCM}
is $\mathcal{O}(1/n)$ where $n$ is the number of iterations. Also,
the optimality gap depends on $K^{2}$. This means that \algref{alg:CBCM}
requires a large number of iterations to return a highly accurate
solution. Thus, in the following we present a variant of the block
coordinate maximization method which we refer to as the greedy block
coordinate maximization (GBCM). This method is based on the Gauss-Southwell
rule which has been numerically shown to achieve a good convergence
rate in practice \cite{Dhillon2011}. The proposed GBCM is described
in \algref{alg:GBCM}. The notation $\left[\mathbf{X}\right]^{+}$
indicates the projection of a Hermitian matrix $\mathbf{X}$ onto the
positive semidefinite cone. In each iteration of \algref{alg:GBCM}
we compute the partial gradient of $\mathcal{L}\bigl(\mu,\bar{\mathbf{S}}^{(n)}\bigr)$
for each $\bar{\mathbf{S}}_{i}$, denoted by $\nabla_{i}\mathcal{L}\bigl(\mu,\bar{\mathbf{S}}^{(n)}\bigr)$
(cf. \eqref{eq:partgraddual}), and use the step size $1/\lambda_{\max}^{2}\bigl(\mathbf{H}_{i}\mathbf{H}_{i}\herm\bigr)$
to move along this direction. As we show in Appendix \ref{subsec:Proof-of-TheoremCBCM},
$\lambda_{\max}^{2}\bigl(\mathbf{H}_{i}\mathbf{H}_{i}\herm\bigr)$
is an upper bound of $\mathcal{L}\bigl(\mu,\bar{\mathbf{S}}^{(n)}\bigr)$
and thus the step size of $1/\lambda_{\max}^{2}\bigl(\mathbf{H}_{i}\mathbf{H}_{i}\herm\bigr)$
always increases the objective. The resulting point is projected to
the positive semidefinite cone, which is then used to compute the
corresponding step length. Among all users, we select one who has
the maximum step length and optimize the covariance matrix of that
user. We remark that, compared to \algref{alg:CBCM}, \algref{alg:GBCM}
only updates the covariance matrix of one user in each iteration.
In our numerical experiments, \algref{alg:GBCM} is shown to achieve
a better convergence rate and thus is used to solve \eqref{eq:dualprob}.\setcounter{algocf}{0} 
\begin{algorithm}[t]
{\small\SetKwInput{KwData}{Initialization}
\SetAlgoNoLine
\DontPrintSemicolon
\LinesNumbered
\renewcommand{\thealgocf}{\arabic{algocf}*} 

\KwIn{$\bar{\mathbf{S}}^{(0)}$}

Set $n\leftarrow0$\;

\Repeat{a stopping criterion is met}{

$k=\underset{1\leq i\leq K}{\argmax}\ \bigl\Vert\bar{\mathbf{S}}_{i}^{(n)}-\bigl[\bar{\mathbf{S}}_{i}^{(n)}+\frac{1}{\lambda_{\max}^{2}\bigl(\mathbf{H}_{i}\mathbf{H}_{i}\herm\bigr)}\nabla_{i}\mathcal{L}\bigl(\mu,\bar{\mathbf{S}}^{(n)}\bigr)\bigr]^{+}\bigr\Vert$\;

Compute $\bar{\mathbf{S}}_{k}^{(n+1)}$ according to \eqref{eq:Sk_opt}
for the chosen $k$

$n\leftarrow n+1$\;

}

\KwOut{$\bar{\mathbf{S}}^{\star}=\bar{\mathbf{S}}^{(n)}$}\caption{$\bar{\mathbf{S}}^{\star}\leftarrow\mathtt{GBCM}(\bar{\mathbf{S}}^{(0)})$\label{alg:GBCM}}
}
\end{algorithm}

\renewcommand{\thealgocf}{\arabic{algocf}} Let $\bar{\mathbf{S}}^{\star}=(\mathbf{\bar{\mathbf{S}}}_{k}^{\star})_{k=1}^{K}$
be the optimal solution of \eqref{eq:dual} for a given $\mu$. Then,
the dual problem is
\begin{equation}
\minimize\ \{g(\mu)\;\bigl|\;\mu\geq0\}.\label{eq:dualprob}
\end{equation}
Since $P-\sum_{k=1}^{K}\tr\bigl(\bar{\mathbf{S}}_{k}^{\star}\bigr)$
is a subgradient of $g(\mu)$, the dual problem in \eqref{eq:dualprob}
can be efficiently solved by a bisection search. In particular, we
increase $\mu_{\min}$ if $P-\sum_{k=1}^{K}\tr\bigl(\bar{\mathbf{S}}_{k}^{\star}\bigr)<0$
and decrease $\mu_{\max}$ otherwise. In summary, the method for solving
\eqref{eq:MIMO:MAC:fixtheta} is outlined in \algref{alg:DD:fixedtheta}.
\begin{algorithm}[t]
{\small\caption{Dual decomposition for solving \eqref{eq:MIMO:MAC:fixtheta}.\label{alg:DD:fixedtheta}}

\SetAlgoNoLine
\DontPrintSemicolon
\LinesNumbered

\KwIn{ $\mu_{\min}=0$, $\mu_{\max}>0$, $\epsilon>0$: desired
accuracy, $\bar{\mathbf{S}}^{(0)}$ }

Set $i\leftarrow0$ \;

\Repeat{$\mu_{\max}-\mu_{\min}<\epsilon$ }{

Set $\mu=\frac{\mu_{\max}+\mu_{\min}}{2}$

Call \algref{alg:GBCM} to obtain $\bar{\mathbf{S}}^{(i+1)}\leftarrow\mathtt{GBCM}(\bar{\mathbf{S}}^{(i)})$

\lIf{$P<\tr\bigl(\bar{\mathbf{S}}^{(i+1)}\bigr)$ }{ Set $\mu_{\min}=\mu$
}\lElse{ Set $\mu_{\max}=\mu$}

$i\leftarrow i+1$\;

}

}
\end{algorithm}

In \cite{Yu:SumCapacity:MIMO_BC:Decomposition:2006}, the authors
proposed a heuristic way to find an appropriate value for $\mu_{\max}$.
We now analytically derive a possible upper limit for the bisection
search in \algref{alg:DD:fixedtheta} as follows. From the \ac{KKT}
condition of \eqref{eq:Sbarmax} we have
\begin{gather*}
\mathbf{H}_{k}\bigl(\bar{\mathbf{H}}_{k}^{-1/2}\bigr)\herm\bigl(\mathbf{I}+\bar{\mathbf{H}}_{k}^{-1/2}\mathbf{H}_{k}\herm\bar{\mathbf{S}}_{k}\mathbf{H}_{k}\bar{\mathbf{H}}_{k}^{-1/2}\bigr)^{-1}\\
\times\bar{\mathbf{H}}_{k}^{-1/2}\mathbf{H}_{k}\herm+\mathbf{M}_{k}=\mu\mathbf{I}
\end{gather*}
where $\mathbf{M}_{k}\succeq\mathbf{0}$ is the Lagrangian multiplier
of the constraints $\bar{\mathbf{S}}_{k}\succeq\mathbf{0}$. Further,
this yields
\begin{gather*}
\mathbf{H}_{k}\bigl(\bar{\mathbf{H}}_{k}^{-1/2}\bigr)\herm\bigl(\mathbf{I}+\bar{\mathbf{H}}_{k}^{-1/2}\mathbf{H}_{k}\herm\bar{\mathbf{S}}_{k}\mathbf{H}_{k}\bar{\mathbf{H}}_{k}^{-1/2}\bigr)^{-1}\bar{\mathbf{H}}_{k}^{-1/2}\mathbf{H}_{k}\herm\bar{\mathbf{S}}_{k}=\mu\bar{\mathbf{S}}_{k}
\end{gather*}
and thus
\begin{gather}
\tr\bigl(\bigl(\mathbf{I}+\bar{\mathbf{H}}_{k}^{-1/2}\mathbf{H}_{k}\herm\bar{\mathbf{S}}_{k}\mathbf{H}_{k}\bar{\mathbf{H}}_{k}^{-1/2}\bigr)^{-1}\bar{\mathbf{H}}_{k}^{-1/2}\mathbf{H}_{k}\herm\nonumber \\
\times\bar{\mathbf{S}}_{k}\mathbf{H}_{k}\bigl(\bar{\mathbf{H}}_{k}^{-1/2}\bigr)\herm\bigr)=\mu\tr\bigl(\bar{\mathbf{S}}_{k}\bigr).
\end{gather}
Note that $\tr\bigl(\bigl(\mathbf{I}+\mathbf{A}\bigr)^{-1}\mathbf{A}\bigr)=\tr\bigl(\bigl(\mathbf{I}+\mathbf{A}^{-1}\bigr)^{-1}\bigr)\leq N_{t}$
and thus the above equality implies $\mu\tr\bigl(\bar{\mathbf{S}}_{k}\bigr)\leq N_{t}$.
Combining this inequality for all users, we have $\mu\leq KN_{t}/P$.
Hence, setting $\mu_{max}=KN_{t}/P$ in \algref{alg:DD:fixedtheta}
guarantees finding the optimal solution to \eqref{eq:MIMO:MAC:fixtheta}.
We note that this upper limit for $\mu$ was not available in \cite{Yu:SumCapacity:MIMO_BC:Decomposition:2006}.

\subsubsection{RIS Optimization}

The RIS optimization is based on the closed-form solution in \cite{zhang2019capacity}.
Specifically, for fixed $\{\bar{\mathbf{S}}_{k}\}_{k=1}^{K}$ and
$\{\theta_{m},m\neq l\}_{m=1}^{N_{s}N_{\mathrm{ris}}}$, the optimization
problem in \eqref{eq:MIMO:MAC:sumrate} with respect to $\theta_{l}$
can be explicitly written as\begin{subequations}\label{eq:MIMO:MAC:fixcovar}
\begin{align}
\underset{\theta_{l}}{\maximize} & \quad\ln\Bigl|\mathbf{I}+\sum\limits _{k=1}^{K}\mathbf{H}_{k}\herm\bar{\mathbf{S}}_{k}\mathbf{H}_{k}\Bigl|\\
\st & \quad\left|\theta_{l}\right|=1.\label{eq:phaseshift}
\end{align}
\end{subequations}To proceed further, we rewrite the objective of
\eqref{eq:MIMO:MAC:fixcovar} as $\log_{2}\Bigl|\mathbf{A}_{l}+\theta_{l}\mathbf{B}_{l}+\theta_{l}^{\ast}\mathbf{B}_{l}\herm\Bigr|$,
where
\begin{gather}
\mathbf{A}_{l}=\mathbf{I}+\sum\limits _{k=1}^{K}\bigl(\mathbf{D}_{k}\herm+\sum\limits _{\underset{m\neq l}{m=1}}^{N_{s}N_{\mathrm{ris}}}\theta_{m}^{*}\mathbf{u}_{m}\herm\mathbf{g}_{k,m}\herm)\bar{\mathbf{S}}_{k}\nonumber \\
\times\bigl(\mathbf{D}_{k}+\sum\limits _{\underset{n\neq l}{n=1}}^{N_{s}N_{\mathrm{ris}}}\theta_{n}\mathbf{g}_{k,n}\mathbf{u}_{n})+\sum\limits _{k=1}^{K}\mathbf{u}_{l}\herm\mathbf{g}_{k,l}\herm\bar{\mathbf{S}}_{k}\mathbf{g}_{k,l}\mathbf{u}_{l},\label{eq:Equ_Al}
\end{gather}
\begin{equation}
\mathbf{B}_{l}=\sum\limits _{k=1}^{K}\bigl(\mathbf{D}_{k}\herm+\sum\limits _{\underset{m\neq l}{m=1}}^{N_{s}N_{\mathrm{ris}}}\theta_{m}^{*}\mathbf{u}_{m}\herm\mathbf{g}_{k,m}\herm\bigr)\bar{\mathbf{S}}_{k}\mathbf{g}_{k,l}\mathbf{u}_{l},\label{eq:Equ_Bl}
\end{equation}
$\mathbf{U}=[\mathbf{u}_{1}\trans\;\mathbf{u}_{2}\trans\;\cdots\;\mathbf{u}_{N_{s}N_{\mathrm{ris}}}\trans]\trans$
and $\mathbf{G}_{k}=[\mathbf{g}_{k,1}\;\mathbf{g}_{k,2}\;\cdots\;\mathbf{g}_{k,N_{s}N_{\mathrm{ris}}}]$.

The optimal solution to \eqref{eq:MIMO:MAC:fixcovar} is then given
by \cite{zhang2019capacity}
\begin{equation}
\theta_{l}^{\star}=\exp(-j\arg(\sigma_{l})),\label{eq:opttheta}
\end{equation}
where $\sigma_{l}$ is the only non-zero eigenvalue of $\mathbf{A}_{l}^{-1}\mathbf{B}_{l}$.
To compute $\sigma_{l}$, a natural way is to calculate $\mathbf{A}_{l}$
and $\mathbf{B}_{l}$ explicitly, and then find the maximum eigenvalue
of $\mathbf{A}_{l}^{-1}\mathbf{B}_{l}$. We now present a more efficient
way to achieve the same goal. First, we can compute a temporary vector
$\mathbf{b}_{l}=\sum\limits _{k=1}^{K}\bigl(\mathbf{D}_{k}\herm+\sum\limits _{\underset{m\neq l}{m=1}}^{N_{\mathrm{ris}}}\theta_{m}^{*}\mathbf{u}_{m}\herm\mathbf{g}_{k,m}\herm\bigr)\bar{\mathbf{S}}_{k}\mathbf{g}_{k,l}$.
Then, $\mathbf{B}_{l}$ can be expressed as $\mathbf{B}_{l}=\mathbf{b}_{l}\mathbf{u}_{l}$
and thus $\mathbf{A}_{l}^{-1}\mathbf{B}_{l}=\mathbf{A}_{l}^{-1}\text{\ensuremath{\mathbf{b}_{l}\mathbf{u}_{l}}}$.
Thanks to this rewriting, it is easy to see that the only non-zero eigenvalue of $\mathbf{A}_{l}^{-1}\mathbf{B}_{l}$
is $\mathbf{u}_{l}\mathbf{A}_{l}^{-1}\mathbf{b}_{l}$. In practice, we
do not need to compute $\mathbf{A}_{l}^{-1}$ explicitly since the
term $\mathbf{A}_{l}^{-1}\mathbf{b}_{l}$ is the solution
to the linear system $\mathbf{A}_{l}\mathbf{x}=\mathbf{b}$. We remark
that this efficient implementation is not available in \cite{zhang2019capacity}.

In summary, the description of the proposed AO algorithm is given
in \algref{alg:AO}. At first, we compute the optimal covariance matrices
for all users, $\{\bar{\mathbf{S}}_{k}\}_{k=1}^{K}$. Next, we sequentially
optimization steps constitute one iteration of \algref{alg:AO}. It
is apparent that each iteration of the AO algorithm increases the achievable
sum-rate. Also, the solution in each iteration of the AO method is
unique and the feasible set is compact. Thus, the convergence of the
AO method to a stationary solution is guaranteed. However, since the
problem \eqref{eq:MIMO:BS:sumrate} is non-convex,
the global optimality of the obtained solution cannot be ensured.
\begin{algorithm}[t]
{\small\caption{AO algorithm for solving \eqref{eq:MIMO:MAC:sumrate}. \label{alg:AO}}

\SetAlgoNoLine
\DontPrintSemicolon
\LinesNumbered 

\KwIn{$\boldsymbol{\theta}^{(0)}\in\varTheta$, $\bar{\mathbf{S}}^{(0)}\in\mathcal{S}$,$n\leftarrow0$}

\Repeat{convergence }{

Set $\bar{\mathbf{S}}^{(n+1)}=\underset{\bar{\mathbf{S}}\in\mathcal{S}}{\arg\max}\ f(\boldsymbol{\theta}^{(n)},\bar{\mathbf{S}})$
using \textbf{Algorithm \ref{alg:DD:fixedtheta}}\;

\For{$l=1,2,\ldots,N_{s}N_{\mathrm{ris}}$}{

$\theta_{l}^{(n+1)}=\exp(-j\arg(\sigma_{l}))$ using \eqref{eq:opttheta}

}

$n\leftarrow n+1$\;

}

}
\end{algorithm}

\subsection{Approximate AO \label{subsec:AO-Grad}}

Although each step in \algref{alg:AO} proceeds via a closed-form
solution, it may take considerable time to return a solution in practice
when the number of RIS elements is large, since \eqref{eq:MIMO:MAC:fixcovar}
needs to be sequentially solved for each RIS element. To make
the aforementioned optimization more efficient, we propose the \emph{approximate}
AO algorithm, in which we improve \algref{alg:AO} by considering
an approximation when optimizing $\boldsymbol{\theta}$ for a given
$\bar{\mathbf{S}}$. More specifically, $\boldsymbol{\theta}^{(n+1)}$
is found as
\begin{align}
\boldsymbol{\theta}^{(n+1)} & =\underset{\boldsymbol{\theta}\in\varTheta}{\arg\max}\ Q_{\mu}(\boldsymbol{\theta},\bar{\mathbf{S}};\boldsymbol{\theta}^{(n)})\triangleq f(\boldsymbol{\theta}^{(n)},\bar{\mathbf{S}})\nonumber \\
 & +\bigl\langle\nabla_{\boldsymbol{\theta}}f\bigl(\boldsymbol{\theta}^{(n)},\bar{\mathbf{S}}),\boldsymbol{\theta}-\boldsymbol{\theta}^{(n)}\bigr\rangle-\frac{1}{\mu}\bigl\Vert\boldsymbol{\theta}-\boldsymbol{\theta}^{(n)}\bigr\Vert^{2}\label{eq:QLB}
\end{align}
where $\left\langle \mathbf{x},\mathbf{y}\right\rangle =2\Re(\mathbf{x}\trans\mathbf{y})$.
Note that the right-hand side $Q_{\mu}(\boldsymbol{\theta},\bar{\mathbf{S}};\boldsymbol{\theta}^{(n)})$
is a quadratic model of $f(\boldsymbol{\theta},\bar{\mathbf{S}})$
around $\boldsymbol{\theta}^{(n)}$ for $\mu>0$. We need to find
$\mu$ such that $Q_{\mu}(\boldsymbol{\theta},\bar{\mathbf{S}};\boldsymbol{\theta}^{(n)})$
becomes a lower bound of $f(\boldsymbol{\theta},\bar{\mathbf{S}})$.
In this regard, let $L_{\boldsymbol{\theta}}(\bar{\mathbf{S}})>0$
be a Lipschitz constant for $\nabla_{\boldsymbol{\theta}}f(\boldsymbol{\theta},\bar{\mathbf{S}})$
for a given $\bar{\mathbf{S}}$, i.e., $\bigl\Vert\nabla_{\boldsymbol{\theta}}f(\boldsymbol{\theta},\bar{\mathbf{S}})-\nabla_{\boldsymbol{\theta}}f(\boldsymbol{\theta}^{\prime},\bar{\mathbf{S}})\bigr\Vert\leq L_{\boldsymbol{\theta}}(\bar{\mathbf{S}})\bigl\Vert\boldsymbol{\theta}-\boldsymbol{\theta}^{\prime}\bigr\Vert,\,\thinspace\forall\boldsymbol{\theta},\boldsymbol{\theta}^{\prime}\in\varTheta$.
Then the following inequality holds
\begin{equation}
f(\boldsymbol{\theta},\bar{\mathbf{S}})\geq Q_{\mu}(\boldsymbol{\theta},\bar{\mathbf{S}};\boldsymbol{\theta}^{(n)}),\forall\boldsymbol{\theta}\in\varTheta\label{eq:ascentLemma}
\end{equation}
for all $\mu\leq\frac{1}{L_{\boldsymbol{\theta}}(\bar{\mathbf{S}})}$.
This above result is in fact an extension of \cite[Lemma 2.1]{beck2009fast}
to complex-valued variables and its proof is given in Appendix \ref{subsec:Proof:AscentLemma}.
It is easy to see that \eqref{eq:QLB} is equivalent~to
\begin{subequations}
\begin{align}
\boldsymbol{\theta}^{(n+1)} & =\underset{\boldsymbol{\theta}\in\varTheta}{\arg\min}\ \bigl\Vert\boldsymbol{\theta}-\bigl(\boldsymbol{\theta}^{(n)}+\mu\nabla_{\boldsymbol{\theta}}f\bigl(\boldsymbol{\theta}^{(n)},\bar{\mathbf{S}})\bigr)\bigr\Vert^{2}\\
 & =P_{\varTheta}\bigl(\boldsymbol{\theta}^{(n)}+\mu\nabla_{\boldsymbol{\theta}}f\bigl(\boldsymbol{\theta}^{(n)},\bar{\mathbf{S}}\bigr)\bigr).
\end{align}
\end{subequations}
Since the Lipschitz constant of $\nabla_{\boldsymbol{\theta}}f(\boldsymbol{\theta},\bar{\mathbf{S}})$,
$L_{\boldsymbol{\theta}}(\bar{\mathbf{S}})$, is not easy to find,
$\mu$ is normally found by a backtracking line search in practice.
The proposed approximate AO is described in \algref{alg:AO:inexact}.

The gradient $\nabla_{\boldsymbol{\theta}}f\bigl(\boldsymbol{\theta},\bar{\mathbf{S}})$
and the projection $P_{\Theta}(\mathbf{\boldsymbol{\theta}})$ needed
to implement \algref{alg:AO:inexact} are given next.\setcounter{thm}{0}
\begin{lem}
The complex gradient of $f\bigl(\boldsymbol{\theta},\bar{\mathbf{S}}\bigr)$
with respect to \textup{$\boldsymbol{\theta}^{*}$ is given by\label{lem:Cov-gradient-1}}
\begin{equation}
\!\nabla_{\boldsymbol{\theta}}f\bigl(\boldsymbol{\theta},\bar{\mathbf{S}})=\vect_{d}\bigl(\sum_{k=1}^{K}\mathbf{G}_{k}\herm\bar{\mathbf{S}}_{k}\mathbf{H}_{k}\bigl(\mathbf{I}+\sum_{n=1}^{K}\mathbf{H}_{n}\herm\bar{\mathbf{S}}_{n}\mathbf{H}_{n}\bigr)^{-1}\mathbf{U}\herm\bigr).\label{eq:Theta_grad}
\end{equation}
\end{lem}
\begin{IEEEproof}
See Lemma 1 in \cite{perovic2020achievable}.
\end{IEEEproof}
The constraint $\bigl|\theta_{l}\bigr|=1$ states that $\theta_{l}$
lies on the unit circle in the complex plane. Thus, for a given point $\boldsymbol{\theta}\in\mathbb{C}^{N_{s}N_{\mathrm{ris}}\times1}$,
$\tilde{\boldsymbol{\theta}}=P_{\Theta}(\mathbf{\boldsymbol{\theta}})$
is given by\footnote{{In practical phase shift models, the amplitude of
$\theta_{l}$ is not necessarily independent of the phase of $\theta_{l}$.
In this case the projection of $\theta_{l}$ onto the set of
feasible reflection coefficients is performed by finding the
feasible reflection coefficient that is closest, based on the
Frobenius norm, to $\theta_{l}$. The same projection can be used
for $\theta_{l}$ computed in \eqref{eq:opttheta}, so that practical
phase shift models can be handled by using \algref{alg:AO:inexact}}.} 
\begin{equation}
\tilde{\theta_{l}}=\begin{cases}
\frac{\theta_{l}}{|\theta_{l}|} & \theta_{l}\neq0\\
e^{j\phi},\phi\in[0,2\pi] & \theta_{l}=0
\end{cases},\;l=1,2,\dots,N_{s}N_{\mathrm{ris}}.\label{eq:projectthetha}
\end{equation}
In particular, $\tilde{\theta_{l}}$ can be any point on the unit
circle if $\theta_{l}=0$, and thus $P_{\Theta}(\mathbf{\boldsymbol{\theta}})$
is not unique.
\begin{algorithm}[t]
{\small\caption{Approximate AO algorithm for solving \eqref{eq:MIMO:MAC:sumrate}.
\label{alg:AO:inexact}}

\SetAlgoNoLine
\DontPrintSemicolon
\LinesNumbered 

\KwIn{$\boldsymbol{\theta}^{(0)}\in\varTheta$, $\bar{\mathbf{S}}^{(0)}\in\mathcal{S}$,
$n\leftarrow0$, $\mu_{0}>0$, $\rho<1$}

\Repeat{convergence }{

Set $\bar{\mathbf{S}}^{(n+1)}=\underset{\bar{\mathbf{S}}\in\mathcal{S}}{\arg\max}\ f(\boldsymbol{\theta}^{(n)},\bar{\mathbf{S}})$
using \textbf{Algorithm \ref{alg:DD:fixedtheta}}\;

\Repeat(\tcc*[f]{line search}){$f(\boldsymbol{\theta}^{(n+1)},\bar{\mathbf{S}}^{(n+1})\geq Q_{\mu_{n}}(\boldsymbol{\theta}^{(n+1)},\bar{\mathbf{S}}^{(n+1)};\boldsymbol{\theta}^{(n)})$}{

$\boldsymbol{\theta}^{(n+1)}=P_{\varTheta}\bigl(\boldsymbol{\theta}^{(n)}+\mu_{n}\nabla_{\boldsymbol{\theta}}f\bigl(\boldsymbol{\theta}^{(n)},\bar{\mathbf{S}}^{(n+1)}\bigr)\bigr)$\;

\If{$f(\boldsymbol{\theta}^{(n+1)},\bar{\mathbf{S}}^{(n+1)})<Q_{\mu_{n}}(\boldsymbol{\theta}^{(n+1)},\bar{\mathbf{S}}^{(n+1)};\boldsymbol{\theta}^{(n)})$
}{$\mu_{n}\leftarrow\rho\mu_{n}$\;}

}

$n\leftarrow n+1$\;

}

}
\end{algorithm}

We remark that the RIS phase shifts are on the unit circle in the
complex plane, which is a manifold. This suggests the possibility
of using the Riemann gradient for the RIS phase shift optimization,
as was done in some previous publications (e.g., \cite{yu2019miso}).
More specifically, the Riemann gradient is given~by
\begin{equation}
\widehat{\nabla}_{\boldsymbol{\theta}}f\bigl(\boldsymbol{\theta},\bar{\mathbf{S}})=\nabla_{\boldsymbol{\theta}}f\bigl(\boldsymbol{\theta},\bar{\mathbf{S}})-\Re\bigl(\nabla_{\boldsymbol{\theta}}f\bigl(\boldsymbol{\theta},\bar{\mathbf{S}})^{\ast}\odot\boldsymbol{\theta}\bigr)\odot\boldsymbol{\theta}
\end{equation}
which is obtained by projecting the Euclidean gradient $\nabla_{\boldsymbol{\theta}}f\bigl(\boldsymbol{\theta},\bar{\mathbf{S}})$
onto the tangent space of the complex unit circle. However, our numerical
experiments have shown that the use of the Riemann gradient method
has no advantage over the Euclidean gradient. Therefore, we
adopt the Euclidean gradient throughout this paper.

\subsection{Alternating Projected Gradient Method (APGM)\label{subsec:PGM-1}}

The main drawback of \textbf{Algorithms \ref{alg:AO}} and \textbf{\ref{alg:AO:inexact}}
is that they rely on \algref{alg:DD:fixedtheta} to solve the optimization
of the covariance matrices when the phase shifts are fixed. Since
\algref{alg:DD:fixedtheta} is a combination of a bisection procedure
and a BCM optimization of $\{\bar{\boldsymbol{S}}_{k}\}$, it may
not be numerically efficient when the number of users is large. Motivated
by the projected gradient step in the $\boldsymbol{\theta}$-update,
in the following we also consider a projected gradient step for the
optimization of the covariance matrices. Specifically, $\bar{\mathbf{S}}^{(n+1)}$
is found as
\begin{equation}
\bar{\mathbf{S}}^{(n+1)}=P_{\mathcal{S}}\bigl(\bar{\mathbf{S}}^{(n)}+\bar{\mu}\nabla_{\bar{\mathbf{S}}}f\bigl(\boldsymbol{\theta}^{(n)},\bar{\mathbf{S}}^{(n)})\bigr)\label{eq:gradientstep_S}
\end{equation}
where $\bar{\mu}$ is the step size for the projected gradient with
respect to $\bar{\mathbf{S}}$. In the above equation, the notation
$\nabla_{\bar{\mathbf{S}}}f\bigl(\boldsymbol{\theta}^{(n)},\bar{\mathbf{S}}^{(n)})$
stands for $\bigl(\nabla_{\bar{\mathbf{S}}_{k}}f\bigl(\boldsymbol{\theta}^{(n)},\bar{\mathbf{S}}^{(n)})\bigr)_{k=1}^{K}$
where $\nabla_{\bar{\mathbf{S}}_{k}}f\bigl(\boldsymbol{\theta},\bar{\mathbf{S}})$
is given by \cite{perovic2020achievable}
\begin{equation}
\nabla_{\bar{\mathbf{S}}_{k}}f\bigl(\boldsymbol{\theta},\bar{\mathbf{S}})=\mathbf{H}_{k}\bigl(\mathbf{I}+\sum\limits _{m=1}^{K}\mathbf{H}_{m}\herm\bar{\mathbf{S}}_{m}\mathbf{H}_{m}\bigr)^{-1}\mathbf{H}_{k}\herm.\label{eq:Skbar_grad}
\end{equation}

The projection of a given point $\bar{\mathbf{S}}$ onto $\mathcal{S}$
admits a water-filling solution as follows. First, $P_{\mathcal{S}}(\bar{\mathbf{S}})$
is explicitly written~as
\begin{equation}
\begin{array}{rl}
\underset{\tilde{\mathbf{S}}_{k}\succeq\mathbf{0}}{\minimize} & ~{\textstyle \bigl\Vert\tilde{\mathbf{S}}-\bar{\mathbf{S}}\bigr\Vert}^{2}=\sum\limits_{k=1}^{K}\bigl\Vert\tilde{\mathbf{S}}_{k}-\bar{\mathbf{S}}_{k}\bigr\Vert^{2}\\
\st & ~{\textstyle \sum\limits_{k=1}^{K}\tr(\tilde{\mathbf{S}}_{k})=P.}
\end{array}\label{eq:projection-1}
\end{equation}
Let $\mathbf{V}_{k}\bar{\mathbf{E}}_{k}\mathbf{V}_{k}^{\dagger}=\bar{\mathbf{S}}_{k}$
be the EVD of $\bar{\mathbf{S}}_{k}$, where $\mathbf{V}_{k}$ is
unitary and $\bar{\mathbf{E}}_{k}$ is diagonal. Then we can write
$\tilde{\mathbf{S}}_{k}=\mathbf{V}_{k}\tilde{\mathbf{E}}_{k}\mathbf{V}_{k}^{\dagger}$
for some $\tilde{\mathbf{E}}_{k}\succeq\mathbf{0}$. Since $\mathbf{V}_{k}$
is unitary, it holds that $\tr(\bar{\mathbf{S}}_{k})=\tr(\bar{\mathbf{E}}_{k})$
and $\tr(\tilde{\mathbf{S}}_{k})=\tr(\tilde{\mathbf{E}}_{k})$, and
hence $\bigl\Vert\tilde{\mathbf{S}}_{k}-\bar{\mathbf{S}}_{k}\bigr\Vert=\bigl\Vert\tilde{\mathbf{E}}_{k}-\bar{\mathbf{E}}_{k}\bigr\Vert$.
That is to say, \eqref{eq:projection-1} is equivalent to
\begin{equation}
\begin{array}{rl}
\underset{\tilde{\mathbf{E}}_{k}\succeq\mathbf{0}}{\minimize} & ~{\textstyle \sum\limits_{k=1}^{K}||\tilde{\mathbf{E}}_{k}-\bar{\mathbf{E}}_{k}||^{2}}\\
\st & ~{\textstyle \sum\limits_{k=1}^{K}\tr(\tilde{\mathbf{E}}_{k})=P.}
\end{array}\label{eq:projection:1-1}
\end{equation}
By direct inspection, we evince that $\tilde{\mathbf{E}}_{k}$
must be diagonal to minimize the objective of \eqref{eq:projection:1-1}.
Let us define $\bar{\mathbf{E}}_{k}=\diag(\bar{\mathbf{e}}_{k})$, $\tilde{\mathbf{E}}_{k}=\diag(\tilde{\mathbf{e}}_{k})$,
$\bar{\mathbf{e}}=[\bar{\mathbf{e}}_{1}\trans,\bar{\mathbf{e}}_{2}\trans,\ldots,\bar{\mathbf{e}}_{K}\trans]\trans$,
and $\tilde{\mathbf{e}}=[\tilde{\mathbf{e}}_{1}\trans,\tilde{\mathbf{e}}_{2}\trans,\ldots,\tilde{\mathbf{e}}_{K}\trans]\trans$.
Then, \eqref{eq:projection:1-1} is reduced to
\begin{equation}
\begin{array}{rl}
\underset{\tilde{\mathbf{e}}\geq0}{\minimize} & ~{\textstyle ||\tilde{\mathbf{e}}-\bar{\mathbf{e}}||^{2}}\\
\st & ~{\textstyle \mathbf{1}_{M}\tilde{\mathbf{e}}=P}
\end{array}\label{eq:projection:compact-1}
\end{equation}
where $M=\sum_{k=1}^{K}n_{k}$ and $\mathbf{1}_{M}$ is the all-ones
vector of length~$M$. We can see that the problem in \eqref{eq:projection:compact-1} admits the following
water-filling~solution
\begin{equation}
\tilde{\mathbf{e}}_{k}=\bigl[\bar{\mathbf{e}}_{k}-\eta\bigr]_{+}
\end{equation}
where $\eta$ is the solution to the equation
\begin{equation}
\mathbf{1}_{M}\bigl[\bar{\mathbf{e}}_{k}-\eta\bigr]_{+}=P,
\end{equation}
which can be found by bisection. More efficient algorithms to find
$\eta$, which use sorting of the entries of the vectors $\bar{\mathbf{e}}_{k}$,
are presented in \cite{Condat2016}. This leads to an algorithm that is referred to as the \emph{APGM} and which is summarized in \algref{alg:APG:adapmomen}.
The term $\bar{Q}_{\mu_{n}}(\boldsymbol{\theta}^{(n)},\bar{\mathbf{S}};\bar{\mathbf{S}}^{(n)})$
in \algref{alg:APG:adapmomen} is the quadratic approximation of $f(\boldsymbol{\theta}^{(n)},\bar{\mathbf{S}})$
around $\bar{\mathbf{S}}^{(n)}$ which is defined as
\begin{align}
\bar{Q}_{\bar{\mu}}(\boldsymbol{\theta},\bar{\mathbf{S}};\bar{\mathbf{S}}^{(n)}) & =f(\boldsymbol{\theta},\bar{\mathbf{S}}^{(n)})\nonumber \\
 & +\sum\limits _{k=1}^{K}\tr\bigl(\bigl(\nabla_{\bar{\mathbf{S}}_{k}}f\bigl(\boldsymbol{\theta},\bar{\mathbf{S}}^{(n)})\bigr)\bigl(\bar{\mathbf{S}}_{k}-\bar{\mathbf{S}}_{k}^{(n)}\bigr)\bigr)\nonumber \\
 & \quad-\frac{1}{2\bar{\mu}}\sum\limits _{k=1}^{K}\bigl\Vert\bar{\mathbf{S}}_{k}-\bar{\mathbf{S}}_{k}^{(n)}\bigr\Vert^{2}.
\end{align}
Accordingly, the projected gradient step in \eqref{eq:gradientstep_S}
is equivalent to $\bar{\mathbf{S}}^{(n+1)}=\argmin\ \{\bar{Q}_{\bar{\mu}}(\boldsymbol{\theta},\bar{\mathbf{S}};\bar{\mathbf{S}}^{(n)})\:|\:\bar{\mathbf{S}}\in\mathcal{S}\}$.
Again, a proper step size $\bar{\mu}$ is chosen such that 
\begin{equation}
f(\boldsymbol{\theta},\bar{\mathbf{S}})\geq\bar{Q}_{\bar{\mu}}(\boldsymbol{\theta},\bar{\mathbf{S}};\bar{\mathbf{S}}^{(n)}).
\end{equation}
In \algref{alg:APG:adapmomen}, an appropriate value for $\bar{\mu}$
is also found by backtracking line search. Since $\nabla_{\bar{\mathbf{S}}}f\bigl(\boldsymbol{\theta},\bar{\mathbf{S}})$
is Lipschitz continuous, the back tracking line search has a finite
number of steps, i.e., when $\bar{\mu}\leq1/L_{\bar{\mathbf{S}}}(\boldsymbol{\theta})$
where $L_{\bar{\mathbf{S}}}(\boldsymbol{\theta})$ is the Lipschitz
constant of $\nabla_{\bar{\mathbf{S}}}f\bigl(\boldsymbol{\theta},\bar{\mathbf{S}})$
for a given $\boldsymbol{\theta}$.
\begin{algorithm}[t]
{\small\caption{APGM algorithm for solving \eqref{eq:MIMO:MAC:sumrate}. \label{alg:APG:adapmomen}}

\SetAlgoNoLine
\DontPrintSemicolon
\LinesNumbered 

\KwIn{$\boldsymbol{\theta}^{(0)}\in\varTheta$, $\bar{\mathbf{S}}^{(0)}\in\mathcal{S}$,
$\mu_{0}>0$, $\bar{\mu}_{0}>0$, $n\leftarrow0$, $\rho<1$.}

\Repeat{convergence}{

\Repeat(\tcc*[f]{line search for $\bar{\mathbf{S}}$}){$f(\boldsymbol{\theta}^{(n)},\bar{\mathbf{S}}^{(n+1)})\geq\bar{Q}_{\bar{\mu}_{n}}(\boldsymbol{\theta}^{(n)},\bar{\mathbf{S}}^{(n+1)};\bar{\mathbf{S}}^{(n)})$}{

$\bar{\mathbf{S}}^{(n+1)}=P_{\varTheta}\bigl(\bar{\mathbf{S}}^{(n)}+\bar{\mu}_{n}\nabla_{\boldsymbol{\theta}}f\bigl(\boldsymbol{\theta}^{(n)},\bar{\mathbf{S}}^{(n)}\bigr)\bigr)$\;\label{alg:APGM:pgS}

\If{$f(\boldsymbol{\theta}^{(n)},\bar{\mathbf{S}}^{(n+1)})<\bar{Q}_{\bar{\mu}_{n}}(\boldsymbol{\theta}^{(n)},\bar{\mathbf{S}}^{(n+1)};\bar{\mathbf{S}}^{(n)})$
}{$\bar{\mu}_{n}\leftarrow\rho\bar{\mu}_{n}$\;}

}

\Repeat(\tcc*[f]{line search for $\boldsymbol{\theta} $}){$f(\boldsymbol{\theta}^{(n+1)},\bar{\mathbf{S}}^{(n+1)})\geq Q_{\mu_{n}}(\boldsymbol{\theta}^{(n+1)},\bar{\mathbf{S}}^{(n+1)};\boldsymbol{\theta}^{(n)})$}{

$\boldsymbol{\theta}^{(n+1)}=P_{\varTheta}\bigl(\boldsymbol{\theta}^{(n)}+\mu_{n}\nabla_{\boldsymbol{\theta}}f\bigl(\boldsymbol{\theta}^{(n)},\bar{\mathbf{S}}^{(n+1)}\bigr)\bigr)$\;

\If{$f(\boldsymbol{\theta}^{(n+1)},\bar{\mathbf{S}}^{(n+1)})<Q_{\mu_{n}}(\boldsymbol{\theta}^{(n+1)},\bar{\mathbf{S}}^{(n+1)};\boldsymbol{\theta}^{(n)})$
}{$\mu_{n}\leftarrow\rho\mu_{n}$\;}

}

$n\leftarrow n+1$\;

}}
\end{algorithm}

The proposed line search procedure ensures that the objective sequence
strictly decreases after each iteration. The detailed convergence
analysis of the APGM can be found in Appendix \ref{sec:APGM_Conv}.
Thus, the APGM algorithm is guaranteed to converge to a stationary point of
\eqref{eq:MIMO:MAC:sumrate}, which is, however, not necessarily a
globally optimal solution.\vspace{-1em}

\subsection{Important Remarks on the Proposed Algorithms\label{subsec:Important-Remarks-on}}

In this section, we explain the novelty of the considered methods
compared to the existing literature and the reasons for proposing
three different optimization methods. First, to efficiently solve
the nonconvex optimization problem \eqref{eq:MIMO:BS:sumrate} we
convert it to \eqref{eq:MIMO:MAC:sumrate} which is convex and more
tractable. Also, the size of each $S_{k}$ in the BC is $N_{t}\times N_{t}$
and the size of $\bar{S}_{k}$ in the dual MAC is $n_{k}\times n_{k}$,
so solving \eqref{eq:MIMO:MAC:sumrate} certainly requires lower complexity
as $N_{t}>n_{k}$.

{Second, we prove in Theorem 1 that
the convergence rate of \algref{alg:CBCM} becomes slow when $K$
is large, which is a new and important result. This motivates us to
consider \algref{alg:GBCM} which can speed up the convergence rate
and also has lower complexity.}

Third, we realize that optimizing each phase shift sequentially admits
a closed-form solution as done in \algref{alg:AO} and is numerically
efficient for a small number of reflecting elements. When the number
of reflecting elements is large, however, it may not lead to an efficient
solution, since many iterations are required. Due to this, we derive
\algref{alg:AO:inexact} where all reflecting elements are simultaneously
optimized by a projected gradient step. In this regard, we have found
that the Riemann gradient, which is more complex, has no advantages
over the Euclidean gradient adopted in this paper.

Fourth, following the same motivation for developing \algref{alg:AO:inexact},
 we optimize all input covariances simultaneously in \algref{alg:APG:adapmomen}.
This algorithm differs from our previous work in \cite{perovic2020achievable},
which is dedicated to single-user MIMO and the covariance
matrix and the reflecting elements are optimized simultaneously. To
make the method in \cite{perovic2020achievable} converge fast, a
scaling step is required in \cite{perovic2020achievable} due to the
different dynamic ranges of the covariance matrix and the phase shifts.
We have found that it is not practically efficient to perform a similar
scaling step for multi-user MIMO due to the significantly different
ranges of the different channels (due to the varying positions of
the users and the presence of multiple RISs in the system). Thus,
in \algref{alg:APG:adapmomen} we propose to optimize the input covariances
and the phase shifts alternately.

Finally, we propose three different algorithms for solving \eqref{eq:MIMO:MAC:sumrate}
and each of them has its own advantages. \algref{alg:AO} is efficient
for small-scale problems and is parameter-free, i.e., no line search
step or data scaling is required. \algref{alg:AO:inexact} becomes
more efficient if the number of reflecting elements is large. On the
other hand, \algref{alg:APG:adapmomen} is generally the most efficient
in terms of complexity if the number of users and the number of RIS
elements are both large. However, its convergence rate is also sensitive
to the local Lipschitz constant of the gradient with respect to each
optimization variable. From Appendix \ref{subsec:Proof-of-TheoremCBCM},
it follows that an upper bound on the Lipschitz constant of
$\nabla_{\bar{\mathbf{S}}}f\bigl(\boldsymbol{\theta}^{(n)},\bar{\mathbf{S}}^{(n)})$
is $\max_{1\leq k\leq K} \lambda_{\max}^{2}\bigl(\mathbf{H}_{k}\mathbf{H}_{k}\herm\bigr)$.
Thus, if $N_{r}$ increases, the Lipschitz constant of $\nabla_{\bar{\mathbf{S}}}f\bigl(\boldsymbol{\theta}^{(n)},\bar{\mathbf{S}}^{(n)})$
is likely to increase accordingly. As a result, the step size in each
iteration of \algref{alg:APG:adapmomen} is decreased, and thus \algref{alg:APG:adapmomen}
takes more iterations to converge. However, the numerical effectiveness
of the three proposed algorithms can only be seen in practice. More
importantly, the considered optimization problem is nonconvex and the proposed
algorithms are only local optimization methods and thus they can be
trapped in poor-performing local optima. Thus, it could be possible
that one of them may avoid this issue to provide a good solution since
they are derived by different optimization frameworks. Based on
extensive numerical experiments we obtain that the three proposed algorithms
come up with different solutions when $N_{t}<\sum_{k=1}^{K}n_{k}$.
It is likely that the degrees of freedom of the resulting system are
reduced in such cases, making it difficult to find a good solution.

{The considerations above are the main motivation
for us to propose three different optimization methods. Furthermore,
in situations where computing resources are abundantly available (e.g.,
multiple processors), the three proposed optimization methods can
be exploited in a concurrent optimization approach. More specifically,
for a given set of channel realizations, we can run the three algorithms
in parallel, can terminate all of them after a given per-determined
time, and can select the best solution among them. This is practically useful in wireless
communications since any algorithm needs to be executed within the
channel coherent time. Alternatively, if there is an indication that
the three proposed algorithms may produce different performances,
we can let them fully converge and choose the best algorithm. In particular,
if all the proposed methods yield approximately the same performance
this indicates that there is a strong possibility that an optimal
solution is reached in the considered simulated settings.}

\section{Computational Complexity\label{sec:Computational-Complexity}}

In this subsection, the computational complexity of each of the proposed
algorithms is obtained by counting the required number of complex
multiplications. In the following complexity derivations, for ease
of exposition, we assume that each user has the same number of antennas,
i.e., $n_{k}=N_{r}$ for all $k=1,2,\dots,K$. Also, we assume that
the number of RIS elements $N_{\mathrm{ris}}$ is significantly larger
than the number of transmit and receive antennas, $N_{t}$ and $N_{r}$,
respectively.
\begin{table*}[t]
\begin{centering}
\begin{tabular}{ccc}
\hline 
Algorithm &  & \multirow{1}{*}{Computational Complexity}\tabularnewline
\hline 
\hline 
AO &  & $\mathcal{O}(TI(KN_{t}N_{r}^{2}+KN_{t}^{2}N_{r}+KN_{r}^{3})+N_{s}N_{\mathrm{ris}}(KN_{t}N_{r}^{2}+KN_{t}^{2}N_{r}+N_{t}^{3}))$\tabularnewline
\hline 
Approximate AO &  & $\mathcal{O}(TI(KN_{t}N_{r}^{2}+KN_{t}^{2}N_{r}+KN_{r}^{3})+I_{\varTheta}KN_{s}N_{\mathrm{ris}}N_{t}N_{r})$\tabularnewline
\hline 
APGM &  & $\mathcal{O}(I_{S}(KN_{t}N_{r}^{2}+KN_{t}^{2}N_{r}+N_{t}^{3}+K^{2}N_{r}^{2})+I_{\varTheta}KN_{s}N_{\mathrm{ris}}N_{t}N_{r})$\tabularnewline
\hline 
\end{tabular}
\par\end{centering}
\centering{}\caption{Computational complexity of one iteration of the AO, approximate AO
and APGM algorithms.\label{tab:complexity}}
\vspace{-0.7cm}
\end{table*}

The computational complexity for the proposed algorithms is presented
in Table \ref{tab:complexity}. The optimization of the covariance matrices
for the AO and the approximate AO algorithms is performed by a dual decomposition
method in \algref{alg:DD:fixedtheta}, which requires $\mathcal{O}(KN_{s}N_{\mathrm{ris}}N_{t}N_{r}+TI(KN_{t}N_{r}^{2}+KN_{t}^{2}N_{r}+KN_{r}^{3}))$
multiplications, where $T$ is the number of outer iterations (i.e.,
lines 3 to 7) in \algref{alg:DD:fixedtheta} and $I$ is the average
number of iterations of \algref{alg:GBCM}. The complexity of optimizing
the RIS phase shifts for the AO algorithm primarily depends on \eqref{eq:Equ_Al}
and \eqref{eq:Equ_Bl}, and is equal to $\mathcal{O}(N_{s}N_{\mathrm{ris}}(KN_{t}N_{r}^{2}+KN_{t}^{2}N_{r}+N_{t}^{3}))$.
On the other hand, the complexity of optimizing the RIS phase shifts
for the approximate AO and APGM algorithms is $\mathcal{O}(I_{\varTheta}KN_{s}N_{\mathrm{ris}}N_{t}N_{r})$,
where $I_{\varTheta}$ is the number of search steps of the line search
procedure for optimizing $\boldsymbol{\theta}$. Optimizing the users
covariance matrices for the APGM algorithm requires $\mathcal{O}(I_{S}(KN_{t}N_{r}^{2}+KN_{t}^{2}N_{r}+N_{t}^{3}+K^{2}N_{r}^{2}))$
multiplications, where $I_{\varTheta}$ is the number of search steps
of the line search procedure. Detailed derivations of the aforementioned
computational complexities are presented in Appendix~\ref{sec:Derived-Complex}.

{For all three optimization algorithms, we observe
that their complexities increase linearly with the number of RIS elements.
While the AO algorithm requires a fixed complexity to optimize the
RIS phase shifts in each iteration, the approximate AO and APGM algorithms
require a complexity for optimizing the RIS phase shifts that depends
on the number of line search steps $I_{\varTheta}$. }{Noticeably,
the per-iteration complexity of the proposed algorithms increases
linearly with the number of RIS elements, which is practically appealing.}{{}
As for the optimization of the covariance matrix, the complexity of an iteration
of \algref{alg:GBCM} for the AO and approximate AO algorithms is approximately
comparable to that of an iteration of the line search
procedure for the APGM algorithm. Hence, the ratio of the total number
of iterations of \algref{alg:GBCM} ($TI$) and the number of line
search steps ($I_{S}$) determines whether the dual decomposition method
for optimizing the covariance matrices is more computationally efficient
than the gradient-based optimization method or not. Further details
on this complexity comparison are presented in Table \ref{tab:Compl_table}.}
\begin{figure}[t]
\begin{centering}
\subfloat[Average achievable sum-rate versus the number of sub-iterations.]{\includegraphics{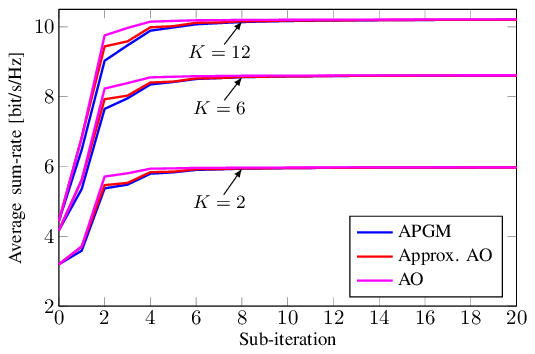}}
\par\end{centering}
\subfloat[Average achievable sum-rate versus the computation time.]{\includegraphics{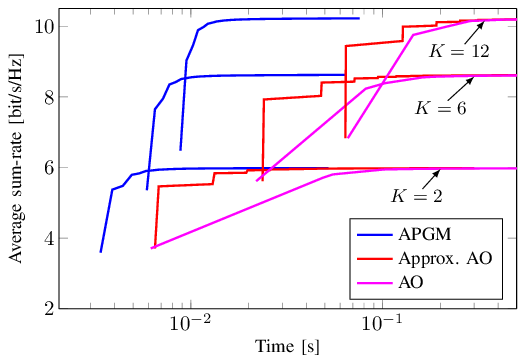}}{\caption{Average achievable sum-rate for the proposed optimization methods
with the direct and RIS-aided links. \label{fig:Rate-DIR-RIS}}
}
\end{figure}
\begin{figure}[t]
\begin{centering}
\subfloat[Average achievable sum-rate versus the number of sub-iterations.]{\includegraphics{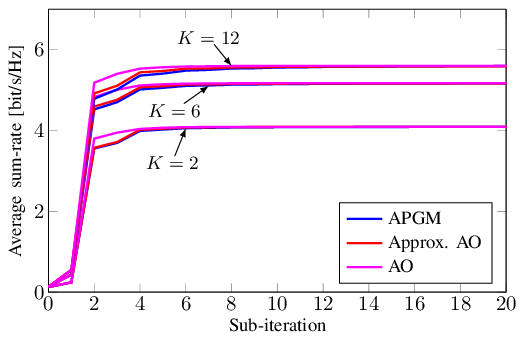}}
\par\end{centering}
\subfloat[Average achievable sum-rate versus the computation time.]{\includegraphics{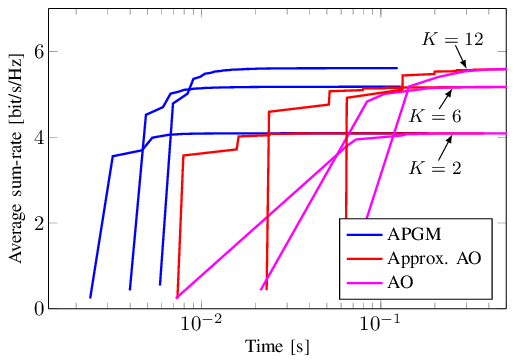}}\caption{Achievable sum-rate for the proposed optimization methods with the RIS-aided
link only.\label{fig:Rate-RIS}}
\end{figure}

\section{Simulation Results\label{sec:Simulation-Results}}

In this section, we evaluate the proposed algorithms in the single-RIS
and multi-RIS setups, with the aid of Monte Carlo simulations. First,
in the single-RIS case, we compare the achievable sum-rates and the
run times of the proposed algorithms. Furthermore, we show the variation
of the achievable sum-rate with the number of transmit antennas at
the BS, with the number of users, and with the number of RIS elements.
We also show the change of the achievable sum-rate with the non-blockage
probability of the direct links. Specifically, we consider the following three different
scenarios: (i) only the direct link (i.e., the first term in
\eqref{eq:Hk_equ}) is present; (ii) only the link via the RIS
(i.e., the second term in \eqref{eq:Hk_equ}) is present; and (iii)
both of these links are present. In the multi-RIS case,
we study the change of the achievable sum-rate with the RIS positions.
More specifically, this study is performed for a constant
number of RIS elements per RIS and for a constant number of RIS elements in the network.

The locations of the BS, the RIS and the users are specified by a
three-dimensional (3D) Cartesian coordinate system. The BS \ac{ULA}
is placed parallel to the \emph{y}-axis and the position of its midpoint
is $(0,l_{t},h_{t})$. The RIS is located in the \emph{xz}-plane
and the position of its midpoint is $(d_{\mathrm{ris}},0,h_{\mathrm{ris}})$.
For simplicity, we assume that all of the users' ULAs are parallel
to the \mbox{\emph{y}-axis} and the midpoint of the \emph{k}-th
user's ULA is $(d_{k},l_{k},h_{k})$. For the considered
system geometry, the distance between the midpoint of the BS ULA and
the midpoint of the RIS is $d_{t,\mathrm{ris}}=\sqrt{d_{\mathrm{ris}}^{2}+l_{t}^{2}+(h_{t}-h_{\mathrm{ris}})^{2}}$,
the distance between the midpoint of the RIS and the midpoint of the
\mbox{\emph{k}-th} user's ULA is $d_{\mathrm{ris},k}=\sqrt{(d_{\mathrm{ris}}-d_{k})^{2}+l_{k}^{2}+(h_{\mathrm{ris}}-h_{k})^{2}}$,
and the distance between the midpoint of the BS ULA and the midpoint
of the \emph{k}-th user's ULA is $d_{t,k}=\sqrt{d_{k}^{2}+(l_{t}-l_{k})^{2}+(h_{t}-h_{k})^{2}}$.

{In the following simulations, all the channel
matrices are modeled according to the Rician fading channel model
with Rician factor equal to 1, as specified in \cite{perovic2020achievable}.
Also, we neglect the spatial correlation among the elements of matrices
$\mathbf{U}$ and $\mathbf{G}_{k}$. The distance-dependent path loss
}for the direct link of the \mbox{\emph{k}-th} user is $\beta_{\mathrm{DIR},k}=(4\pi/\lambda)^{2}d_{t,k}^{\alpha_{\mathrm{DIR}}}$,
where $\alpha_{\mathrm{DIR}}$ denotes the path loss exponent of the
direct link. The far-field \ac{FSPL} for the RIS-aided link of the \emph{k}-th
user $\beta_{\mathrm{RIS},k}$ is equal to $\beta_{\mathrm{RIS},k}^{-1}=G_{t}G_{r}\lambda^{4}\cos\gamma_{t}\cos\gamma_{r}/(256\pi^{2}d_{t,\mathrm{ris}}^{2}d_{\mathrm{ris},k}^{2})$,
where $\gamma_{t}$ is the angle between the propagation direction of the incident wave 
 and the normal to the RIS, and $\gamma_{r}$ is the angle
between the normal to the RIS and the propagation direction of the reflected wave
\cite[Eq. (7), (9)]{tang2020wireless}. Hence, we have $\cos\gamma_{t}=l_{t}/d_{t,\mathrm{ris}}$
and $\cos\gamma_{r}=l_{k}/d_{\mathrm{ris},k}$. Also, $G_{t}$ and
$G_{r}$ represent the transmit and receive antenna gains respectively,
which are both set to 2, since we assume that these antennas
radiate/sense signals to/from the relevant half-space \cite{tang2020wireless}.
Finally, $\sqrt{\beta_{\mathrm{DIR},k}^{-1}/N_{0}}$ and $\sqrt{\beta_{\mathrm{RIS},k}^{-1}/N_{0}}$
are embedded as scaling factors in $\mathbf{D}_{k}$ and $\mathbf{G}_{k}$,
respectively.

As for the simulation setup, the parameters are $f=2\,\mathrm{GHz}$
(i.e., $\lambda=15\,\mathrm{cm}$), $s_{t}=s_{r}=s_{\mathrm{ris}}=\lambda/2=7.5\,\mathrm{cm}$,
$l_{t}=20\,\mathrm{m}$, $h_{t}=10\,\mathrm{m}$, $d_{\mathrm{ris}}=30\,\mathrm{m}$,
$h_{\mathrm{ris}}=5\,\mathrm{m}$, $N_{t}=8$, $\alpha_{\mathrm{DIR}}=3$,
$P=1\,\mathrm{W}$, and $N_{0}=-110\thinspace\mathrm{dB}$. The RIS
consists of $N_{\mathrm{ris}}=225$ elements placed in a \mbox{$15\times15$}
square formation. We assume that all users are equipped with $N_{r}=2$
antennas. The users' coordinates are randomly selected such that $d_{k}$
is chosen from a uniform distribution between $200\,\mathrm{m}$ and
$500\,\mathrm{m}$ with a resolution of $2\,\mathrm{m}$, $l_{k}$
is chosen from a uniform distribution between $1\,\mathrm{m}$ and
$70\,\mathrm{m}$ with a resolution of $1\,\mathrm{m}$, and $h_{k}$
is chosen from a uniform distribution between $1.5\,\mathrm{m}$ and
$2\,\mathrm{m}$ with a resolution of $1\,\mathrm{cm}$.
For the dual decomposition optimization methods,
we have $\mu_{\max}=KN_{t}/P$ and $\epsilon=10^{-5}$. For the gradient-based
optimization methods, the initial step size value is 10000. All results
are averaged over 1000 independent channel~realizations.

The achievable sum-rate for the proposed methods when 
the channel consists of the direct and RIS-aided links, and the \mbox{RIS-aided}
link only, are shown in Figs. \ref{fig:Rate-DIR-RIS} and \ref{fig:Rate-RIS},
respectively. We assume that each iteration of \textbf{Algorithms
\ref{alg:AO}}, \textbf{\ref{alg:AO:inexact}} and \textbf{\ref{alg:APG:adapmomen}}
consists of two \emph{sub-iterations}, so that in each \emph{sub-iteration}
we optimize all the users' covariance matrices or all the RIS
phase shifts. Hence, the achievable sum-rate in Figs. \ref{fig:Rate-DIR-RIS}
and \ref{fig:Rate-RIS} is computed after every sub-iteration.{{}
For presenting the achievable sum-rate versus the rum time of the algorithms,
 we neglect the initial achievable sum-rate which is the same
for all the algorithms and present first the achievable sum-rate obtained
after the first sub-iteration. All the algorithms have the same
initial values of the users' covariance matrices and the initial values of the RIS phase
shifts that are randomly generated for each channel realization. Also,
all of the algorithms are executed on the same laptop computer (4
core processors with a frequency of 1.5 GHz and 16 GB RAM). From the figures, 
we see that the proposed optimization algorithms
reach the same objective value which is a locally optimal achievable
sum-rate. In addition, all the algorithms achieve the same objective value for a relatively
low number of iterations. This is particularly visible for the AO
algorithm. However, each iteration for the optimization of the RIS phase shifts
of the AO algorithm is very time intensive because of the sequential optimization
of the RIS phase shifts. In addition, the approximate AO and the APGM algorithms
show approximately the same convergence behavior with respect to the
number of iterations for a system with two users, while for a larger
number of users the approximate AO algorithm has only a slight advantage compared
to the APGM algorithm. On the other hand, the APGM algorithm requires less time to reach
a locally optimal achievable rate. For a larger number of users (e.g.,
$K=12$) the approximate AO algorithm has almost the same convergence time as
the AO algorithm. It seems that the gradient-based optimization of the users'
covariance matrices is generally more time efficient than the GBCM
optimization method, especially in a system with large $K$. }As expected,
the achievable sum-rate in multi-user communications is higher when
the direct link is present, similar to the
achievable rate in point-to-point communications reported
in \cite{perovic2020achievable}. Moreover, the achievable sum-rate
increases with the number of users $K$, and the increase is more
substantial when the direct link is present. It seems that the lack
of capability of adjusting the amplitude of the reflection coefficients
 prevents the RIS from achieving
a significant suppression of the multi-user interference. Therefore,
higher achievable sum-rate gains can be expected when a part of a
signal is transmitted via the direct link, since the BS, thanks to its amplitude
adjustment capabilities, is better equipped to suppress the aforementioned
interference.

In the previous setting, the APGM algorithm is shown to provide the best
performance. However, as explained in Section \ref{subsec:Important-Remarks-on},
the convergence rate of the APGM algorithm depends on the Lipschitz constant
of the gradients. Thus, the APGM algorithm is not universally the best. To illustrate
this point, we consider a slightly different setting, where the number
of receive antennas per user is $N_{r}=4$, the number of RIS elements
is $N_{\mathrm{ris}}=100$ and $d_{k}=100\,\mathrm{m}$ for all the
users. The other parameters are kept to their default values as for Fig.
\ref{fig:Rate-RIS}. Fig. \ref{fig:Ach_rate_except} shows the
achievable sum-rate of the proposed methods when only the RIS-aided link
is present. Compared to Fig. \ref{fig:Rate-RIS}, the number of receive
antennas is larger in Fig. \ref{fig:Ach_rate_except}, which likely
increases the Lipschitz constant of the gradients $\nabla_{\bar{\mathbf{S}}}f\bigl(\boldsymbol{\theta}^{(n)},\bar{\mathbf{S}}^{(n)})$
as explained analytically in Section \ref{subsec:Important-Remarks-on}.
This in turn forces the APGM algorithm to take more iterations to converge.
As can be seen in Fig. \ref{fig:Ach_rate_except}, the AO algorithm needs the
least number of iterations and time to converge. On the other hand,
the APGM algorithm requires more time and more than a magnitude of order of
iterations to return a solution.
\begin{figure}[t]
\begin{centering}
\subfloat[Average achievable{{} sum-rate versus the number of sub-iterations.}]{\includegraphics{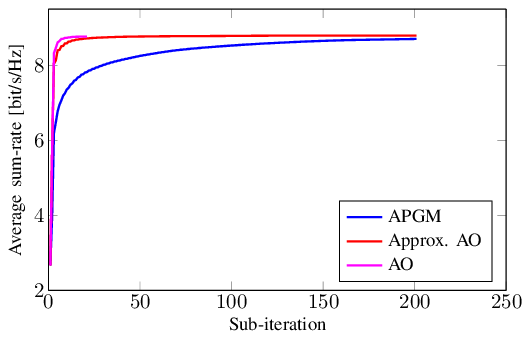}}
\par\end{centering}
\subfloat[Average achievable{{} sum-rate versus the computation time.}]{\includegraphics{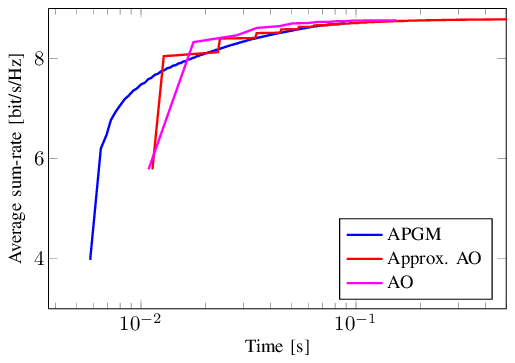}}{\caption{Average achievable sum-rate for the proposed optimization
methods with the RIS-aided link only, and slightly changed simulation parameters
setup compared to Fig. 3 ($K=2$).\label{fig:Ach_rate_except}}
}
\end{figure}

{In the next numerical experiment, we show that the
three proposed optimization algorithms can produce different sum-rate
performance since the considered problem is non-convex. This justifies
our motivation for proposing three algorithms for attaining the best
achievable sum-rate. To this end, we consider a scenario where the
are $3$ users and the number of transmit antennas is $N_{t}=4$.
The users' positions are randomly generated. The other simulation
parameters are the same as those in Fig. \ref{fig:Ach_rate_except}.
In this considered scenario, we remark that the degrees of freedom,
i.e., the multiplexing gain, is very low and upper-bounded to $N_{t}=4$~}\footnote{{The maximum number of degrees of freedom, i.e., the multiplexing
gain, for the considered system is given by $\min(N_{t},KN_{r})=N_{t}.$}}{.}\textbf{{{} }}{Hence,
this multi-user system can transmit up to 4 independent data streams,
which is equal to the maximum number of data streams for an individual
user. Thus, most of the eigenvalues of $(\bar{\mathbf{S}}_{k})_{k=1}^{K}$
are 0 (i.e., the corresponding eigenchannels receive no power) to
maximize the achievable sum-rate. Consequently, it becomes more challenging
to find an optimal solution according to \cite[Appendix A]{luo2008dynamic}.
To illustrate this, we plot in Fig. \ref{fig:Ach_rate_snap_shot}
the average performance of the proposed algorithms over 100 channel
realizations. For each channel realization, we also compute the best
performance of the three algorithms when they are convergent, which
is dubbed ``Best'' in Fig. \ref{fig:Ach_rate_snap_shot}. We notice
that the algorithms achieve different sum-rates. Therefore, in systems
with low degrees of freedom, all three algorithms need to be run to increase
the probability that the maximum sum-rate is obtained. In general,
through a large number of numerical experiments, we have observed
that the APGM algorithm usually provides good trade-offs between the achievable
sum-rate and complexity, and thus it is the recommended choice in
most scenarios.}{}
\begin{figure}[t]
{\includegraphics{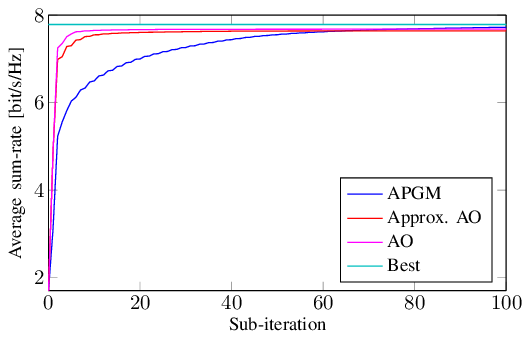}\caption{Average achievable sum-rate of each proposed algorithm
and their best sum-rate performance.\label{fig:Ach_rate_snap_shot}}
}
\end{figure}

\begin{figure}[t]
\includegraphics{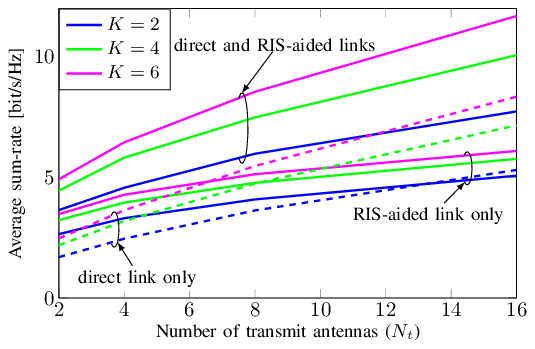}\caption{Achievable sum-rate versus the number of transmit antennas ($N_{t}$).\label{fig:AO_DIR}}
\end{figure}
{In Fig. \ref{fig:AO_DIR}, we show the achievable
sum-rate for the best of the three methods versus the number of transmit
antennas $N_{t}$. The achievable} sum-rate has an approximately
logarithmic shape. Also, it can be observed that the achievable sum-rate
increases with the number of users. However, it seems that this increase
gradually declines with the increase of the number of users. At the
same time, the achievable sum-rate increases with the number of transmit
antennas. For 6 users and 2 transmit antennas, for example, a 99\,\%
increase in the achievable sum-rate is obtained by adding the RIS
to the multi-user system.
\begin{figure}[t]
\includegraphics{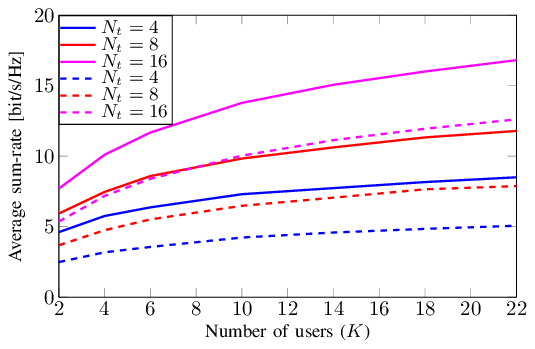}\caption{Average achievable sum-rate versus the number of users ($K$) in the
case of direct and RIS-aided links (solid lines) and direct link
only (dashed lines). \label{fig:RatevsUsers}}
\end{figure}

{In Fig. \ref{fig:RatevsUsers}, we present the achievable sum-rate for the
best of the three proposed methods versus the number of users.
Since }we aim to investigate the increase of the achievable sum-rate
due to the presence of the RIS, we show the rate in the two cases
where the channel consists of the direct and RIS-aided links (solid lines),
and the direct link only (dashed lines). The achievable sum-rate has
an approximately logarithmic shape. In more detail, we observe
that, due to the multi-user diversity gain, the achievable sum-rate
is approximately proportional to $\min(N_{t},KN_{r}).$ This means
that, with the increase of the number of users, the rate becomes proportional
to the number of transmit antennas $N_{t}.$ Hence, the difference
between the achievable sum-rates for two different values of $N_{t}$
increases with the number of users. Also, we see that the presence
of the RIS increases the achievable sum-rate and influences the slope
of each achievable sum-rate. Specifically, the slope of the
achievable rate is slightly larger when the RIS is present,
particularly for small values of $K$. As the aforementioned slope reduces with
$K$, the achievable sum-rate negligibly increases for large values of $K$
when the RIS is present and the same occurs for systems without an
\ac{RIS}, as was already shown in~\cite{Nam:GreedyScheduling:SZFDPC:2010}.

{In	Fig. \ref{fig:DPC_vs_LIN}, we examine the benefits of
DPC over linear precoding in the presence of RIS. Specifically, 
 we plot the sum-rate versus
the number of RIS elements obtained by DPC and by linear precoding.
For linear precoding, we slightly modify the method introduced in
\cite{pan2020intelligent} to find the sum-rate. Also, to understand
how DPC and linear precoding perform in the case of discrete phase
shifts at the RIS, we show the achievable sum-rates for DPC
and linear precoding when the phase shifts are quantized with 1 or
2 bits. For these cases, the sum-rates are achieved by mapping the
continuous phase shifts (when the corresponding algorithm convergences)
to the closest discrete phase shift. Note that for DPC we report the
best performance among the three proposed algorithms. The achievable
sum-rates for DPC and linear precoding increase approximately logarithmically
with the number of RIS elements. Due to the ability of \ac{DPC} to
presubtract the known interference, the proposed algorithms always
provide a larger achievable sum-rate than the algorithm in \cite{pan2020intelligent}.
Moreover, the performance advantage of DPC becomes more noticeable
when the number of RIS elements is large. This may be attributed to
the assumption that the amplitudes of the reflection coefficients
of the RIS are not tunable \cite{Stefan:2021:multiuserRIS}. It can
also be seen that even a low resolution (i.e., 1- or 2-bit resolution)
for the phase shifts of the RIS is sufficient to achieve
performance comparable to the case with continuous phase shifts. }
\begin{figure}[t]
{\includegraphics{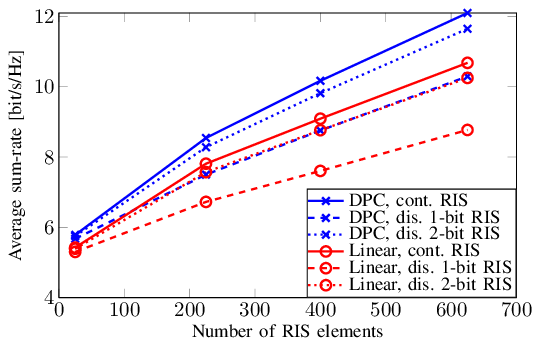}\caption{{Average achievable sum-rates for DPC and linear
precoding versus the number of quantization bits for the phase shifts of the RIS. The parameter setup is the same as for Fig. \ref{fig:Rate-DIR-RIS}
and $K=6$. \label{fig:DPC_vs_LIN}}}
}
\end{figure}

{Since perfect \ac{CSI} is hard to obtain, particularly
in RIS-aided communication systems, the achievable sum-rate subject to
 imperfect \ac{CSI} is shown in Fig. \ref{fig:Rate_imp_CSI}.
We assume that the estimated channel matrix can be represented as
a sum of the true channel matrix and an estimation error matrix, which
consists of i.i.d. elements that are distributed according to $\mathcal{CN}(0,\sigma^{2})$.
Also, it is assumed that imperfect CSI does not influence the \ac{FSPL}.
 We see that the achievable sum-rate exhibits
a moderate decrease even for relatively large values of $\sigma^{2}$. For example,
the achievable sum-rate reduces by 1.3\,bit/s/Hz for $\sigma^{2}=0.9$.
Hence, the proposed optimization algorithms can be efficiently used
in RIS-aided communications even if the CSI knowledge is imperfect. }
\begin{figure}[t]
{\includegraphics{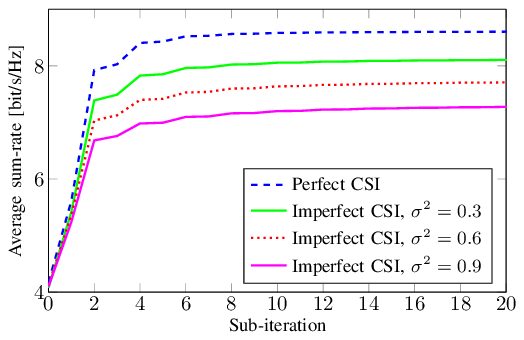}\caption{{Achievable sum-rate assuming perfect and
imperfect CSI. The parameter setup is the same as for Fig. \ref{fig:Rate-DIR-RIS}
and $K=6$. \label{fig:Rate_imp_CSI}}}
}
\end{figure}

\begin{figure}[t]
\includegraphics{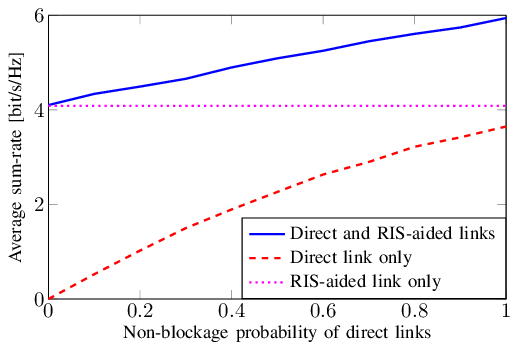}\caption{Average achievable sum-rate versus the non-blockage probability of
the direct links ($K=2$).\label{fig:Rate_vs_prob}}
\end{figure}
In the previous simulations, we considered communication scenarios
when the direct link is present or blocked. However, in most practical
scenarios the direct link is blocked only for a certain fraction of
 time. Hence, to obtain a more comprehensive picture, Fig. \ref{fig:Rate_vs_prob} presents (for the case of 2 users) the achievable
sum-rate versus the ``non-blockage probability'' of the direct links
(i.e., the probability that each direct link is unblocked). Specifically, it is
assumed that the non-blockage probability is the same for all users,
and that the blockage of the direct link occurs independently for
each user \cite{van2021reconfigurable}. When the direct and RIS-aided links
are present, the achievable sum-rate changes linearly with the non-blockage
probability. When the non-blockage probability is equal to one, the achievable
sum-rate boils down to the corresponding rate in Fig. \ref{fig:AO_DIR},
and when the non-blockage probability is equal zero, the achievable
sum-rate is the same as when only the RIS-aided link is considered. If only
the direct link is present, the achievable sum-rate has an approximately
logarithmic shape and its maximum value coincides with the corresponding
rate in Fig. \ref{fig:AO_DIR}.

The per-iteration computational complexities of
the proposed optimization algorithms when the direct link (DL) is
present or blocked are shown in Table \ref{tab:Compl_table}. 
The relevant numbers of iterations of the dual decomposition method $T$
and $I$, and the number of line search steps $I_{S}$ and $I_{\varTheta}$
are averaged over the first five iterations of the proposed algorithms,
since all the algorithms converge within five iterations. As expected,
the AO algorithm has the largest computational complexity which is mainly due
to the sequential optimization of the RIS phase shifts. The approximate
AO algorithm achieves a lower complexity than the AO algorithm, due to the more efficient
gradient-based optimization of the RIS phase shifts. For a large number
of users, the number of line search steps $I_{S}$ for the APGM algorithm is
significantly lower than the total number of iterations $TI$ of \algref{alg:GBCM}
for the AO and approximate AO algorithms. Therefore, the APGM algorithm is particularly
suitable for application to systems with a large number of users.

\begin{table}[t]
\centering{}\resizebox{\columnwidth}{!}{{\scriptsize{}}%
\begin{tabular}{|c|c|ccc|cccc|ccc|}
\hline 
\multirow{2}{*}{DL} & \multirow{2}{*}{$K$} & \multicolumn{3}{c|}{AO} & \multicolumn{4}{c|}{{Approximate AO}} & \multicolumn{3}{c|}{APGM}\tabularnewline
\cline{3-12} \cline{4-12} \cline{5-12} \cline{6-12} \cline{7-12} \cline{8-12} \cline{9-12} \cline{10-12} \cline{11-12} \cline{12-12} 
 &  & $T$ & $I$ & $C_{\mathrm{AO}}$ & $T$ & $I$ & $I_{\varTheta}$ & $C_{\mathrm{A-AO}}$ & $I_{S}$ & $I_{\varTheta}$ & $C_{\mathrm{APGM}}$\tabularnewline
\hline 
\multirow{3}{*}{\begin{turn}{90}
Present\,
\end{turn}} & 2 & 24 & 3 & 211392 & 21 & 3 & 1 & 28368 & 4 & 1 & 10592\tabularnewline
\cline{2-12} \cline{3-12} \cline{4-12} \cline{5-12} \cline{6-12} \cline{7-12} \cline{8-12} \cline{9-12} \cline{10-12} \cline{11-12} \cline{12-12} 
 & 6 & 26 & 8 & 540864 & 23 & 8 & 1 & 207072 & 4 & 1 & 28064\tabularnewline
\cline{2-12} \cline{3-12} \cline{4-12} \cline{5-12} \cline{6-12} \cline{7-12} \cline{8-12} \cline{9-12} \cline{10-12} \cline{11-12} \cline{12-12} 
 & 12 & 27 & 14 & 1309248 & 24 & 14 & 1 & 720576 & 5 & 1 & 58240\tabularnewline
\hline 
\multirow{3}{*}{\begin{turn}{90}
Blocked\,
\end{turn}} & 2 & 24 & 3 & 211392 & 21 & 3 & 1 & 28368 & 3 & 1 & 9744\tabularnewline
\cline{2-12} \cline{3-12} \cline{4-12} \cline{5-12} \cline{6-12} \cline{7-12} \cline{8-12} \cline{9-12} \cline{10-12} \cline{11-12} \cline{12-12} 
 & 6 & 26 & 7 & 514656 & 23 & 7 & 1 & 183888 & 3 & 1 & 26448\tabularnewline
\cline{2-12} \cline{3-12} \cline{4-12} \cline{5-12} \cline{6-12} \cline{7-12} \cline{8-12} \cline{9-12} \cline{10-12} \cline{11-12} \cline{12-12} 
 & 12 & 27 & 13 & 1254816 & 24 & 13 & 1 & 672192 & 3 & 2 & 95424\tabularnewline
\hline 
\end{tabular}{\scriptsize{}}}\caption{{Comparison of the per-iteration computational complexities
of the AO, approximate AO and APGM algorithms ($N_{t}=8$, $N_{r}=2$,
$N_{\mathrm{ris}}=225$). \label{tab:Compl_table}}}
\end{table}

{In the rest of this section, we analyze the performance
of the proposed algorithms in the multi-RIS case. The simulation setup
has the same parameters as in the single-RIS case except that the
BS coordinate $l_{t}$ is 30\,m; }the {\emph{$x$}}{-coordinates
$d_{k}$ are chosen from a uniform distribution between $275\,\mathrm{m}$
and $325\,\mathrm{m}$ with a resolution of $1\,\mathrm{m}$, and
$l_{k}$ is chosen from a uniform distribution between $5\,\mathrm{m}$
and $55\,\mathrm{m}$ with a resolution of $1\,\mathrm{m}$ }\footnote{{In order to fully understand the influence of the
positions of the RISs, the users are randomly placed over a smaller area of size
$50\times50\,\mathrm{m}$.}}{. The midpoint of the BS is aligned with the center
of the user-populated area, and the distance between them is $D=300\,\mathrm{m}$.
Besides the aforementioned RIS, which is denoted as RIS 1, three more
RISs are added in the considered communication system. They are denoted
as RIS 2, RIS 3 and RIS 4, and the positions of their midpoints are
located at $(D-d_{\mathrm{ris}},0,5\thinspace\mathrm{m})$, $(d_{\mathrm{ris}},60\thinspace\mathrm{m},5\thinspace\mathrm{m})$
and $(D-d_{\mathrm{ris}},60\thinspace\mathrm{m},5\thinspace\mathrm{m})$,
respectively}\footnote{{It is worth noting that RIS 1 and RIS 3 are always
at the same distance from the midpoint of the BS, and that RIS 2 and
RIS 4 are always at the same distance from the center of the user-populated
area.}}. The achievable sum-rate for the considered system
versus $d_{\mathrm{ris}}$ is presented in Fig. \ref{fig:Rates-vs-dist}.
As expected, the largest achievable rate is obtained when the RISs
are located close to the BS and to the user-populated area. Interestingly,
the achievable sum-rate shows only a very modest increase when RIS
1 and RIS 2 simultaneously reflect the incoming signals, as compared to the case when
only one of these two surfaces performs reflection, although the number
of RIS elements is doubled (i.e., it is now equal to 450). A
15\,\% increase in the achievable sum-rate is obtained for the RISs placed
close to the center and an increase of around 35\,\% for the RISs placed in the
vicinity of the BS and the user-populated area is obtained. This confirms
the fact that the best placements for the RISs are close to the transmitter
or the receiver. Similar trends can be observed when all four RISs
are simultaneously reflecting the incoming signals and the total number
of RIS elements is 900. 
\begin{figure}[t]{
{\includegraphics{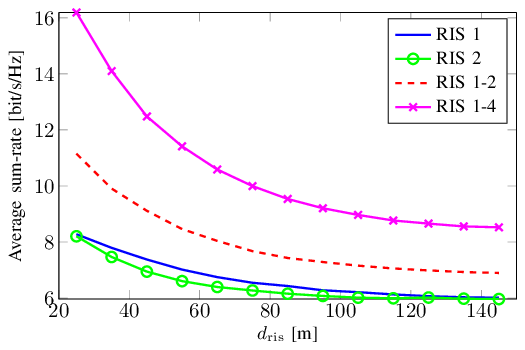}\caption{Average achievable sum-rates versus $d_{\mathrm{ris}}$
($K=6$).\label{fig:Rates-vs-dist}}}
}
\end{figure}

{Lastly, we study the achievable sum-rate for the
multi-RIS case where the total number of RIS elements in the system
is constant. More precisely, all of the RIS elements can be located
on a single \ac{RIS} or equally distributed among a subset of the
\acp{RIS}. The achievable sum-rate when the total number of RIS elements
is equal to 400 is shown in Fig. \ref{fig:Rates-const-Nris}. Using more \acp{RIS}
for signal transmission is only beneficial for small values of $d_{\mathrm{ris}}$,
i.e., when the \acp{RIS} are in the vicinity of the BS and the user-populated
area. On the other hand, the single-RIS transmission provides superior
rates for larger values of $d_{\mathrm{ris}}$, i.e., when the RISs are located
far from the BS and the user-populated area. It seems that placing
all of the reflecting elements on a single RIS can provide a higher
array gain which is essential for signal transmission via weak communication
links (i.e., when the RISs are located far from the BS and the user-populated
area). }
\begin{figure}[t]
{\includegraphics{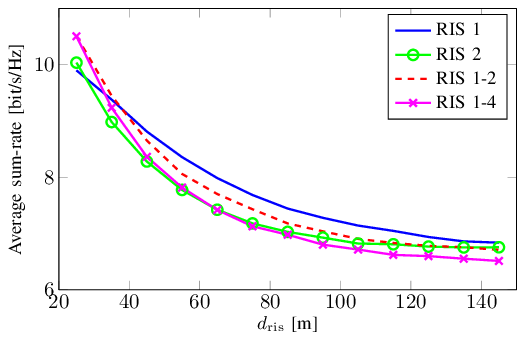}\caption{{Average achievable sum-rates for a constant total
number of RIS elements ($K=6$).\label{fig:Rates-const-Nris}}}
}
\end{figure}

\section{Conclusion\label{sec:Conclusion}}

In this paper, we exploited the well-known \ac{BC}-\ac{MAC} duality
for the achievable sum-rate optimization in a multi-user \ac{BC}
in the presence of one or multiple RISs. Due to the non-convexity of the
considered optimization problem, we proposed three algorithms that
provide the same achievable sum-rate. In the proposed algorithms,
the users\textquoteright{} covariance matrices were optimized by a
dual decomposition method with a \ac{BCM} or a gradient-based method,
while the optimal RIS phase shifts were sequentially computed by using
a closed-form expression or in parallel by a gradient-based
method. Also, we presented a computation complexity analysis for the
proposed algorithms. Simulation results showed that
the proposed algorithms usually achieve the same sum-rate. However,
they can produce different sum-rates for some specific situations,
due to the non-convexity of the considered problem. Also, the gradient-based
optimization methods were generally more time efficient, particularly
when the number of RIS elements is large. Furthermore, we demonstrated
that the proposed algorithms are easily implementable in the multi-RIS case as well,
and that can provide significant achievable sum-rate gains that depend
on the placement of \ac{RIS} elements in a BC.

\appendices{}

\section{Computational Complexities \label{sec:Derived-Complex}}

The complexity of the AO algorithm is determined by the computation of the covariance
matrices $\{\bar{\mathbf{S}}_{k}\}_{k=1}^{K}$ and the RIS phase shifts
$\{\theta_{m}\}_{m=1}^{N_{s}N_{\mathrm{ris}}}$ in \algref{alg:AO}.
 At first, we need to compute all the users' channel matrices.
To compute $\mathbf{F}(\boldsymbol{\theta})\mathbf{U}$ requires $N_{\mathrm{ris}}N_{t}$
multiplications and it is common for all users. To obtain all $\mathbf{G}_{k}\mathbf{F}(\boldsymbol{\theta})\mathbf{U}$ matrices,
we need $\mathcal{O}(KN_{s}N_{\mathrm{ris}}N_{t}N_{r})$ multiplications.
To reduce the complexity of computing $\bar{\mathbf{H}}_{k}$, instead
of implementing \eqref{eq:Hbar_k} directly, we compute and store
$\mathbf{H}_{\mathrm{sum}}=\mathbf{I}+\sum_{j=1}^{K}\mathbf{H}_{j}\herm\bar{\mathbf{S}}_{j}\mathbf{H}_{j}$,
which requires $\mathcal{O}(KN_{t}N_{r}^{2}+KN_{t}^{2}N_{r})$ multiplications.
In each iteration of \algref{alg:GBCM}, we first compute each gradient
$\nabla_{i}\mathcal{L}\bigl(\mu,\bar{\mathbf{S}}^{(n)}\bigr)$ according
to \eqref{eq:partgraddual}. The matrix inversion $\mathbf{H}_{\mathrm{sum}}^{-1}$
is executed only once per \algref{alg:GBCM} iteration, so its complexity
is neglected. The product $\mathbf{H}_{k}\mathbf{H}_{\mathrm{sum}}^{-1}\mathbf{H}_{k}\herm$
needs $\mathcal{O}(N_{t}N_{r}^{2}+N_{t}^{2}N_{r})$ multiplications.
The complexity of obtaining all $\lambda_{\max}^{2}\bigl(\mathbf{H}_{k}\mathbf{H}_{k}\herm\bigr)$
terms can be neglected since they can be calculated once and used in every
iteration of \algref{alg:GBCM}. The projection onto the semidefinite
cone requires $\mathcal{O}(N_{r}^{3})$ multiplications and the complexity
of line 3 of \algref{alg:GBCM} can be approximated by $\mathcal{O}(KN_{t}N_{r}^{2}+KN_{t}^{2}N_{r}+KN_{r}^{3})$.
The computation of $\bar{\mathbf{H}}_{k}^{-1}=(\mathbf{H}_{\mathrm{sum}}-\mathbf{H}_{k}\herm\bar{\mathbf{S}}_{k}\mathbf{H}_{k})^{-1}$
has a complexity of $\mathcal{O}(N_{t}^{3})$ and the computation of $\mathbf{H}_{k}\bar{\mathbf{H}}_{k}^{-1}\mathbf{H}_{k}\herm$
has a complexity of $\mathcal{O}(N_{t}N_{r}^{2}+N_{t}^{2}N_{r})$.
The \ac{EVD} of $\mathbf{H}_{k}\bar{\mathbf{H}}_{k}^{-1}\mathbf{H}_{k}\herm$
requires $\mathcal{O}(N_{r}^{3})$ multiplications, while the complexity
of computing $\bar{\mathbf{S}}_{k}^{(n+1)}$ is $\mathcal{O}(N_{r}^{3})$.
Finally, the complexity of one iteration of \algref{alg:GBCM} is
$\mathcal{O}(KN_{t}N_{r}^{2}+KN_{t}^{2}N_{r}+KN_{r}^{3})$. As a
result, the computational complexity of one iteration of \algref{alg:DD:fixedtheta}
is $\mathcal{O}(KN_{s}N_{\mathrm{ris}}N_{t}N_{r}+TI(KN_{t}N_{r}^{2}+KN_{t}^{2}N_{r}+KN_{r}^{3}))$,
where $T$ is the required number of outer iterations (i.e., lines
3 to 7) in \algref{alg:DD:fixedtheta}, and $I$ is the average number
of iterations of \algref{alg:GBCM}. In our case, $T$ is the smallest
integer that satisfies the inequality $\mu_{\max}/2^{T}<\epsilon$.

The complexity of computing the optimal RIS phase shifts is primarily
dependent on \eqref{eq:Equ_Al} and \eqref{eq:Equ_Bl}. Let us define
$\mathbf{C}_{k}=\mathbf{H}_{k}-\theta_{l}\mathbf{g}_{k,l}\mathbf{u}_{l}$
to simplify the derivation. The complexity of computing
the matrix $\mathbf{C}_{k}\bar{\mathbf{S}}_{k}\mathbf{C}_{k}$
is $\mathcal{O}(N_{t}N_{r}^{2}+N_{t}^{2}N_{r})$. In a similar manner,
the complexity of computing $\mathbf{u}_{l}\herm\mathbf{g}_{k,l}\herm\bar{\mathbf{S}}_{k}\mathbf{g}_{k,l}\mathbf{u}_{l}$
is equal to $\mathcal{O}(N_{t}N_{r}^{2}+N_{t}^{2}N_{r})$. Hence,
the complexity of computing $\mathbf{A}_{l}$ in \eqref{eq:Equ_Al}
is $\mathcal{O}(KN_{t}N_{r}^{2}+KN_{t}^{2}N_{r})$. Also, we need
$\mathcal{O}(KN_{t}^{2}N_{r})$ more multiplications to obtain $\mathbf{B}_{l}$
in \eqref{eq:Equ_Bl}. Inverting $\mathbf{A}_{l}$ requires $\mathcal{O}(N_{t}^{3})$
multiplications. The same complexity is required for computing $\mathbf{A}_{l}^{-1}\mathbf{B}_{l}$
and for obtaining the \ac{EVD} of that product. The complexity of
computing a single RIS phase shift is $\mathcal{O}(KN_{t}N_{r}^{2}+KN_{t}^{2}N_{r}+N_{t}^{3})$,
which gives a total~of $\mathcal{O}(N_{s}N_{\mathrm{ris}}(KN_{t}N_{r}^{2}+KN_{t}^{2}N_{r}+N_{t}^{3}))$
complex multiplications for the whole RIS.

In summary, the complexity of one overall iteration (i.e., lines 2
to 6 in \algref{alg:AO}) of the AO algorithm is given by
\begin{align}
C_{\mathrm{AO}}= & \,\mathcal{O}(TI(KN_{t}N_{r}^{2}+KN_{t}^{2}N_{r}+KN_{r}^{3})\nonumber \\
 & \;\;\;+N_{s}N_{\mathrm{ris}}(KN_{t}N_{r}^{2}+KN_{t}^{2}N_{r}+N_{t}^{3})).\label{eq:AO_compl-2}
\end{align}

The complexity of the approximate AO algorithm differs from the complexity of the AO algorithm
in the optimization of the RIS phase shifts. To optimize the RIS phase
shifts of the approximate AO algorithm, we have to calculate the gradient $\nabla_{\boldsymbol{\theta}}f\bigl(\boldsymbol{\theta},\bar{\mathbf{S}})$
in \eqref{eq:Theta_grad}. The computation of $\mathbf{H}_{\mathrm{sum}}^{-1}\mathbf{U}\herm$
requires $\mathcal{O}(N_{s}N_{\mathrm{ris}}N_{t}^{2})$ complex multiplications.
Since the complexity of $\mathbf{G}_{k}\herm\bar{\mathbf{S}}_{k}\mathbf{H}_{k}$
is $\mathcal{O}(N_{s}N_{\mathrm{ris}}N_{t}N_{r})$, the complexity
of $\sum_{k=1}^{K}\mathbf{G}_{k}\herm\bar{\mathbf{S}}_{k}\mathbf{H}_{k}$
is $\mathcal{O}(KN_{s}N_{\mathrm{ris}}N_{t}N_{r})$. In addition,
we need $\mathcal{O}(N_{s}N_{\mathrm{ris}}N_{t}^{2})$ multiplications
to compute the diagonal elements of $\sum_{k=1}^{K}\mathbf{G}_{k}\herm\bar{\mathbf{S}}_{k}\mathbf{H}_{k}\mathbf{H}_{\mathrm{sum}}^{-1}\mathbf{U}\herm$.
As a result, the complexity of computing the gradient $\nabla_{\boldsymbol{\theta}}f\bigl(\boldsymbol{\theta},\bar{\mathbf{S}})$
is $\mathcal{O}(N_{s}N_{\mathrm{ris}}N_{t}^{2}+KN_{s}N_{\mathrm{ris}}N_{t}N_{r})$.
The projection $P_{\varTheta}(\cdot)$ has a negligibly small complexity.
The complexity of the line search procedure assuming $I_{\varTheta}$
search steps is $\mathcal{O}(I_{\varTheta}KN_{s}N_{\mathrm{ris}}N_{t}N_{r})$.
The upper-bound on $I_{\varTheta}$ corresponds to the smallest integer
that satisfies the inequality $\mu_{0}/\rho^{I_{\varTheta}}<L_{\boldsymbol{\theta}}$,
where $\mu_{0}$ is the initial step size and $L_{\boldsymbol{\theta}}$
is the Lipschitz constant of $\nabla_{\boldsymbol{\theta}}f(\boldsymbol{\theta},\bar{\mathbf{S}})$.

Therefore, the complexity of one overall iteration (i.e., lines 2
to 9 in \algref{alg:AO:inexact}) of the approximate AO algorithm
is given by
\begin{align}
C_{\mathrm{A-AO}}= & \,\mathcal{O}(TI(KN_{t}N_{r}^{2}+KN_{t}^{2}N_{r}+KN_{r}^{3})\nonumber \\
 & \qquad+I_{\varTheta}KN_{s}N_{\mathrm{ris}}N_{t}N_{r}).\label{eq:AO_compl-1-1}
\end{align}

The complexity of the APGM algorithm is determined by two gradient optimization
loops: one for the optimization of the covariance matrices (lines 2 to 7 in
\algref{alg:APG:adapmomen}) and the other for the optimization of the RIS phase shifts
(lines 8 to 13 in \algref{alg:APG:adapmomen}). To compute
the gradient $\nabla_{\bar{\mathbf{S}}_{k}}f\bigl(\boldsymbol{\theta},\bar{\mathbf{S}})$
in \eqref{eq:Skbar_grad}, we need to calculate $\mathbf{H}_{k}\mathbf{H}_{\mathrm{sum}}^{-1}\mathbf{H}_{k}\herm$
which requires $\mathcal{O}(N_{t}N_{r}^{2}+N_{t}^{2}N_{r})$ multiplications.
Therefore, the complexity of calculating $\nabla_{\bar{\mathbf{S}}}f\bigl(\boldsymbol{\theta},\bar{\mathbf{S}}\bigr)$
is $\mathcal{O}(KN_{t}N_{r}^{2}+KN_{t}^{2}N_{r})$. Additional $\mathcal{O}(K^{2}N_{r}^{2}+KN_{r}^{3})$
multiplications are required to calculate the projection $P_{\mathcal{S}}(\cdot)$.
To compute $f(\boldsymbol{\theta}^{(n)},\bar{\mathbf{S}}^{(n+1)}),$
we need $\mathcal{O}(KN_{t}N_{r}^{2}+KN_{t}^{2}N_{r}+N_{t}^{3})$
multiplications. The additional complexity for calculating $\bar{Q}_{\bar{\mu}_{n}}(\boldsymbol{\theta}^{(n)},\bar{\mathbf{S}}^{(n+1)};\bar{\mathbf{S}}^{(n)})$
is negligible. Hence, the complexity of the line search procedure
assuming $I_{S}$ search steps is $\mathcal{O}(I_{S}(KN_{t}N_{r}^{2}+KN_{t}^{2}N_{r}+N_{t}^{3}+K^{2}N_{r}^{2}))$,
which is exactly the complexity of optimization loop for the covariance matrices
(lines 2 to 7 in \algref{alg:APG:adapmomen}). The upper-bound
on $I_{S}$ corresponds to the smallest integer that satisfies the inequality $\bar{\mu}_{0}/\rho^{I_{S}}<L_{\bar{\mathbf{S}}}$,
where $\bar{\mu}_{0}$ is the initial step size and $L_{\bar{\mathbf{S}}}$
is the Lipschitz constant of $\nabla_{\bar{\mathbf{S}}}f(\boldsymbol{\theta},\bar{\mathbf{S}})$.
Similar to the approximate AO algorithm, the complexity of the optimization loop for the covariance
matrices (lines 8 to 13 in \algref{alg:APG:adapmomen})
is $\mathcal{O}(I_{\varTheta}KN_{\mathrm{ris}}N_{t}N_{r}),$ where
$I_{\varTheta}$ is the number of search steps.

Therefore, the complexity of one overall iteration (lines 2 to 14
in \algref{alg:APG:adapmomen}) of the APGM algorithm is equal
\begin{align}
C_{\mathrm{APGM}}= & \,\mathcal{O}(I_{S}(KN_{t}N_{r}^{2}+KN_{t}^{2}N_{r}+N_{t}^{3}+K^{2}N_{r}^{2})\nonumber \\
 & \qquad+I_{\varTheta}KN_{s}N_{\mathrm{ris}}N_{t}N_{r}).\label{eq:PGM_compl-1}
\end{align}

\section{Proof of Theorem \ref{thm:CBCM}\label{subsec:Proof-of-TheoremCBCM}}

We define the partial gradient of $\mathcal{L}(\mu,\bar{\mathbf{S}})$
with respect to each component $\bar{\mathbf{S}}_{k}$ as
\begin{subequations}
\label{eq:partgraddual}
\begin{align}
\nabla_{k}\mathcal{L}(\mu,\bar{\mathbf{S}}) & =\nabla_{\bar{\mathbf{S}}_{k}}\mathcal{L}(\mu,\bar{\mathbf{S}})\\
 & =\mathbf{H}_{k}\bigl(\bar{\mathbf{H}}_{k}+\mathbf{H}_{k}\herm\bar{\mathbf{S}}_{k}\mathbf{H}_{k}\bigr)^{-1}\mathbf{H}_{k}\herm-\mu\mathbf{I},
\end{align}
\end{subequations}
where $\bar{\mathbf{H}}_{k}=\mathbf{I}+\sum\limits _{i=1,i\neq k}^{K}\mathbf{H}_{i}\herm\bar{\mathbf{S}}_{i}\mathbf{H}_{i}$.
The next step is to study the Lipschitz constant of $\nabla_{k}\mathcal{L}(\mu,\bar{\mathbf{S}})$.
Towards this end, let $\bar{\mathbf{S}}(\mathbf{X)}_{k}=(\bar{\mathbf{S}}_{1},\ldots\bar{\mathbf{S}}_{k-1},\mathbf{X},\bar{\mathbf{S}}_{k+1},\ldots,\bar{\mathbf{S}}_{K})$
denote the point where the $k$-th component of $\bar{\mathbf{S}}$
is replaced by $\mathbf{X}$. Then the inequalities in \eqref{eq:blockLipschitz},
shown at the top of the next page, hold. Thus, $\lambda_{\max}^{2}\bigl(\mathbf{H}_{k}\mathbf{H}_{k}\herm\bigr)$
is an upper bound for the Lipschitz constant of $\nabla_{k}\mathcal{L}(\mu,\bar{\mathbf{S}})$
and $M=\underset{1\leq k\leq K}{\max}\ \lambda_{\max}^{2}\bigl(\mathbf{H}_{k}\mathbf{H}_{k}\herm\bigr)$
is an upper bound for the Lipschitz constant of $\nabla\mathcal{L}(\mu,\bar{\mathbf{S}})$, which is defined as 
\begin{equation}
\nabla\mathcal{L}(\mu,\bar{\mathbf{S}})=\bigl(\nabla_{1}\mathcal{L}(\mu,\bar{\mathbf{S}}),\nabla_{2}\mathcal{L}(\mu,\bar{\mathbf{S}}),\ldots,\nabla_{K}\mathcal{L}(\mu,\bar{\mathbf{S}})\bigr).
\end{equation}
More specifically, we have
\begin{subequations}
{\small
\begin{gather}
\bigl\Vert\nabla\mathcal{L}(\mu,\bar{\mathbf{S}})-\nabla\mathcal{L}(\mu,\bar{\mathbf{S}}^{\prime})\bigr\Vert\triangleq\sqrt{\sum\limits\limits _{k=1}^{K}\bigl\Vert\nabla_{k}\mathcal{L}(\mu,\bar{\mathbf{S}})-\nabla_{k}\mathcal{L}(\mu,\bar{\mathbf{S}}^{\prime})\bigr\Vert^{2}}\\
\leq\sqrt{\sum\limits\limits _{k=1}^{K}\lambda_{\max}^{4}\bigl(\mathbf{H}_{k}\mathbf{H}_{k}\herm\bigr)\bigl\Vert\bar{\mathbf{S}}_{k}-\bar{\mathbf{S}}_{k}^{\prime}\bigr\Vert^{2}}\\
\leq M\sqrt{\sum\limits\limits _{k=1}^{K}\bigl\Vert\bar{\mathbf{S}}_{k}-\bar{\mathbf{S}}_{k}^{\prime}\bigr\Vert^{2}}=M\bigl\Vert\bar{\mathbf{S}}-\bar{\mathbf{S}}^{\prime}\bigr\Vert.
\end{gather}
}
\end{subequations}
Accordingly, Theorem \ref{thm:CBCM} is a direct result
of \cite[Theorem 2]{Hong2017}.
\begin{figure*}[tbh]
\begin{subequations}
\label{eq:blockLipschitz}
\begin{align}
\bigl\Vert\nabla_{k}\mathcal{L}(\mu,\bar{\mathbf{S}}(\mathbf{X)}_{k})-\nabla_{k}\mathcal{L}(\mu,\bar{\mathbf{S}}(\mathbf{Y)}_{k})\bigr\Vert & =\bigl\Vert\mathbf{H}_{k}\bigl(\bar{\mathbf{H}}_{k}+\mathbf{H}_{k}\herm\mathbf{X}\mathbf{H}_{k}\bigr)^{-1}\mathbf{H}_{k}\herm-\mathbf{H}_{k}\bigl(\bar{\mathbf{H}}_{k}+\mathbf{H}_{k}\herm\mathbf{Y}\mathbf{H}_{k}\bigr)^{-1}\mathbf{H}_{k}\herm\bigr\Vert\\
 & =\bigl\Vert\mathbf{H}_{k}\bigl[\bigl(\bar{\mathbf{H}}_{k}+\mathbf{H}_{k}\herm\mathbf{X}\mathbf{H}_{k}\bigr)^{-1}-\bigl(\bar{\mathbf{H}}_{k}+\mathbf{H}_{k}\herm\mathbf{Y}\mathbf{H}_{k}\bigr)^{-1}\bigr]\mathbf{H}_{k}\herm\bigr\Vert\\
 & =\bigl\Vert\mathbf{H}_{k}\bigl(\bar{\mathbf{H}}_{k}+\mathbf{H}_{k}\herm\mathbf{X}\mathbf{H}_{k}\bigr)^{-1}\mathbf{H}_{k}\herm\bigl(\mathbf{X}-\mathbf{Y}\bigr)\mathbf{H}_{k}\bigl(\bar{\mathbf{H}}_{k}+\mathbf{H}_{k}\herm\mathbf{Y}\mathbf{H}_{k}\bigr)^{-1}\mathbf{H}_{k}\herm\bigr\Vert\\
 & \leq\lambda_{\max}^{2}\bigl(\mathbf{H}_{k}\mathbf{H}_{k}\herm\bigr)\bigl\Vert\mathbf{X}-\mathbf{Y}\bigr\Vert
\end{align}
\end{subequations}
\end{figure*}

\section{Proof of \eqref{eq:ascentLemma} \label{subsec:Proof:AscentLemma}}

To simplify the notations, we write $f(\boldsymbol{\theta})$ instead
of $f(\boldsymbol{\theta},\bar{\mathbf{S}})$ in this appendix. Let
$\tilde{\boldsymbol{\theta}}=\begin{bmatrix}\Re(\boldsymbol{\theta})\trans & \Im(\boldsymbol{\theta})\trans\end{bmatrix}\trans$
and $\tilde{f}(\tilde{\boldsymbol{\theta}})$ be the corresponding
function of real variables, i.e., $\tilde{f}(\tilde{\boldsymbol{\theta}})=f(\boldsymbol{\theta})$.
From the definition of $\nabla_{\boldsymbol{\theta}}f(\boldsymbol{\theta})$,
we have
\begin{align}
\bigl\Vert\nabla_{\tilde{\boldsymbol{\theta}}}f(\tilde{\boldsymbol{\theta}})-\nabla_{\tilde{\boldsymbol{\theta}}}f(\tilde{\boldsymbol{\theta}}^{\prime})\bigr\Vert & =2\bigl\Vert\nabla_{\boldsymbol{\theta}}f(\boldsymbol{\theta})-\nabla_{\boldsymbol{\theta}}f(\boldsymbol{\theta}^{\prime})\bigr\Vert\nonumber \\
 & \leq2L_{\boldsymbol{\theta}}(\bar{\mathbf{S}})\bigl\Vert\boldsymbol{\theta}-\boldsymbol{\theta}^{\prime}\bigr\Vert=2L_{\varTheta}\bigl\Vert\tilde{\boldsymbol{\theta}}-\tilde{\boldsymbol{\theta}}^{\prime}\bigr\Vert
\end{align}
which means that $2L_{\boldsymbol{\theta}}(\bar{\mathbf{S}})$ is
the Lipschitz constant of $\nabla_{\tilde{\boldsymbol{\theta}}}f(\tilde{\boldsymbol{\theta}})$.
From \cite[Lemma 2.1]{beck2009fast}, the following inequality holds
for any \mbox{$\bar{L}\geq L_{\boldsymbol{\theta}}(\bar{\mathbf{S}})$:}
\begin{gather}
f(\boldsymbol{\theta})=\tilde{f}(\tilde{\boldsymbol{\theta}})\geq\tilde{f}(\tilde{\boldsymbol{\theta}}^{(n)})+\nabla_{\tilde{\boldsymbol{\theta}}}\tilde{f}(\tilde{\boldsymbol{\theta}}^{(n)})\trans(\tilde{\boldsymbol{\theta}}-\tilde{\boldsymbol{\theta}}^{(n)})-\frac{\bar{L}}{2}\left\Vert \tilde{\boldsymbol{\theta}}-\tilde{\boldsymbol{\theta}}^{(n)}\right\Vert ^{2}\nonumber \\
=f(\boldsymbol{\theta}^{(n)})+\nabla_{\tilde{\boldsymbol{\theta}}}\tilde{f}(\tilde{\boldsymbol{\theta}}^{(n)})\trans(\tilde{\boldsymbol{\theta}}-\tilde{\boldsymbol{\theta}}^{(n)})-\frac{\bar{L}}{2}\left\Vert \boldsymbol{\theta}-\boldsymbol{\theta}^{(n)}\right\Vert ^{2}.
\end{gather}
The proof is completed by letting $\mu=\frac{2}{\bar{L}}\leq\frac{1}{L_{\boldsymbol{\theta}}(\bar{\mathbf{S}})}$
and noting that $\nabla_{\tilde{\boldsymbol{\theta}}}\tilde{f}(\tilde{\boldsymbol{\theta}}^{(n)})\trans(\tilde{\boldsymbol{\theta}}-\tilde{\boldsymbol{\theta}}^{(n)})=2\Re\bigl(\nabla_{\boldsymbol{\theta}}f(\boldsymbol{\theta}^{(n)})\herm(\boldsymbol{\theta}-\boldsymbol{\theta}^{(n)})\bigr)=\bigl\langle\nabla_{\boldsymbol{\theta}}f(\boldsymbol{\theta}^{(n)}),\boldsymbol{\theta}-\boldsymbol{\theta}^{(n)}\bigr\rangle$.

\section{Convergence Analysis of the APGM Algorithm \label{sec:APGM_Conv}}

In this appendix, we provide the convergence analysis of the proposed
APGM algorithm. Our arguments follows those in \cite{bolte2014proximal}. Since
$L_{\bar{\mathbf{S}}}(\boldsymbol{\theta}^{(n)})$ is the Lipschitz
constant of $\nabla_{\bar{\mathbf{S}}}f(\bar{\mathbf{S}},\boldsymbol{\theta}^{(n)})$,
it follows~that
\begin{multline}
f(\boldsymbol{\theta}^{(n)},\bar{\mathbf{S}}^{(n+1)})\geq f(\boldsymbol{\theta}^{(n)},\bar{\mathbf{S}}^{(n)})\\
+\sum\limits _{k=1}^{K}\tr\bigl(\bigl(\nabla_{\bar{\mathbf{S}}_{k}}f\bigl(\boldsymbol{\theta},\bar{\mathbf{S}}^{(n)})\bigr)\bigl(\bar{\mathbf{S}}_{k}^{(n+1)}-\bar{\mathbf{S}}_{k}^{(n)}\bigr)\bigr)\\
\quad-\frac{L_{\bar{\mathbf{S}}}(\boldsymbol{\theta}^{(n)})}{2}\sum_{k=1}^{K}\bigl\Vert\bar{\mathbf{S}}_{k}^{(n+1)}-\bar{\mathbf{S}}_{k}^{(n)}\bigr\Vert^{2}.\label{eq:descentS}
\end{multline}
The projected gradient step in line \ref{alg:APGM:pgS} of \algref{alg:APG:adapmomen}
implies
\begin{align}
\sum\limits _{k=1}^{K}\tr\bigl(\bigl(\nabla_{\bar{\mathbf{S}}_{k}}f\bigl(\boldsymbol{\theta},\bar{\mathbf{S}}^{(n)})\bigr)\bigl(\bar{\mathbf{S}}_{k}^{(n+1)}-\bar{\mathbf{S}}_{k}^{(n)}\bigr)\bigr)\nonumber \\
-\frac{1}{2\mu_{n}}\sum\limits _{k=1}^{K}\bigl\Vert\bar{\mathbf{S}}_{k}^{(n+1)}-\bar{\mathbf{S}}_{k}^{(n)}\bigr\Vert^{2}\ge & 0.\label{eq:optimalityS}
\end{align}
Combining \eqref{eq:descentS} and \eqref{eq:optimalityS} yields
\begin{gather}
f(\boldsymbol{\theta}^{(n)},\bar{\mathbf{S}}^{(n+1)})\geq f(\boldsymbol{\theta}^{(n)},\bar{\mathbf{S}}^{(n)})\nonumber \\
+\frac{1}{2}\bigl(\frac{1}{\mu_{n}}-L_{\bar{\mathbf{S}}}(\boldsymbol{\theta}^{(n)})\bigr)\sum\limits _{k=1}^{K}\bigl\Vert\bar{\mathbf{S}}_{k}^{(n+1)}-\bar{\mathbf{S}}_{k}^{(n)}\bigr\Vert^{2}.
\end{gather}
Similarly, from the $\boldsymbol{\theta}$-update we have
\begin{multline}
f(\boldsymbol{\theta}^{(n+1)},\bar{\mathbf{S}}^{(n+1)})\geq f(\boldsymbol{\theta}^{(n)},\bar{\mathbf{S}}^{(n+1)})\\
+\frac{1}{2}\bigl(\frac{1}{\bar{\mu}_{n}}-L_{\boldsymbol{\theta}}(\bar{\mathbf{S}}^{(n+1)})\bigr)\bigl\Vert\boldsymbol{\theta}^{(n+1)}-\boldsymbol{\theta}^{(n)}\bigr\Vert^{2}
\end{multline}
and thus
\begin{multline}
f(\boldsymbol{\theta}^{(n+1)},\bar{\mathbf{S}}^{(n+1)})\geq f(\boldsymbol{\theta}^{(n)},\bar{\mathbf{S}}^{(n)})\\
+\frac{1}{2}\bigl(\frac{1}{\mu_{n}}-L_{\bar{\mathbf{S}}}(\boldsymbol{\theta}^{(n)})\bigr)\sum\limits _{k=1}^{K}\bigl\Vert\bar{\mathbf{S}}_{k}^{(n+1)}-\bar{\mathbf{S}}_{k}^{(n)}\bigr\Vert^{2}\\
+\frac{1}{2}\bigl(\frac{1}{\bar{\mu}_{n}}-L_{\boldsymbol{\theta}}(\bar{\mathbf{S}}^{(n+1)})\bigr)\bigl\Vert\boldsymbol{\theta}^{(n+1)}-\boldsymbol{\theta}^{(n)}\bigr\Vert^{2}.
\end{multline}
The backtracking line search procedure ensures that $1/\mu_{n}\geq L_{\bar{\mathbf{S}}}(\boldsymbol{\theta}^{(n)})/\rho$
and $1/\bar{\mu}_{n}\geq L_{\boldsymbol{\theta}}(\bar{\mathbf{S}}^{(n+1)}/\rho$
which results in
\begin{multline}
f(\boldsymbol{\theta}^{(n+1)},\bar{\mathbf{S}}^{(n+1)})\geq f(\boldsymbol{\theta}^{(n)},\bar{\mathbf{S}}^{(n)})\\
+\frac{L_{\bar{\mathbf{S}}}(\boldsymbol{\theta}^{(n)})}{2}\bigl(\frac{1}{\rho}-1\bigr)\sum\limits _{k=1}^{K}\bigl\Vert\bar{\mathbf{S}}_{k}^{(n+1)}-\bar{\mathbf{S}}_{k}^{(n)}\bigr\Vert^{2}\\
+\frac{L_{\boldsymbol{\theta}}(\bar{\mathbf{S}}^{(n+1)})}{2}\bigl(\frac{1}{\rho}-1\bigr)\bigl\Vert\boldsymbol{\theta}^{(n+1)}-\boldsymbol{\theta}^{(n)}\bigr\Vert^{2}.
\label{eq:last}
\end{multline}
Since $\rho<1$, the inequality in \eqref{eq:last} implies that the sequence $\{f(\boldsymbol{\theta}^{(n)},\bar{\mathbf{S}}^{(n)})\}$
is strictly increasing. Moreover, $f(\boldsymbol{\theta}^{(n)},\bar{\mathbf{S}}^{(n)})$
is bounded from above due to the continuity $f(\boldsymbol{\theta},\bar{\mathbf{S}})$
and the compactness of the feasible set. Hence, the objective sequence
$\{f(\boldsymbol{\theta}^{(n)},\bar{\mathbf{S}}^{(n)})\}$ is convergent.
Moreover, according to \cite{bolte2014proximal}, it can be shown that
the sequence $(\boldsymbol{\theta}^{(n)},\bar{\mathbf{S}}^{(n)})$
indeed converges to a stationary point of \eqref{eq:MIMO:MAC:sumrate},
but we skip the details for brevity.

\bibliographystyle{IEEEtran}
\bibliography{IEEEabrv,references,IEEEexample}

\begin{thebibliography}{10}
\providecommand{\url}[1]{#1}
\csname url@samestyle\endcsname
\providecommand{\newblock}{\relax}
\providecommand{\bibinfo}[2]{#2}
\providecommand{\BIBentrySTDinterwordspacing}{\spaceskip=0pt\relax}
\providecommand{\BIBentryALTinterwordstretchfactor}{4}
\providecommand{\BIBentryALTinterwordspacing}{\spaceskip=\fontdimen2\font plus
\BIBentryALTinterwordstretchfactor\fontdimen3\font minus
  \fontdimen4\font\relax}
\providecommand{\BIBforeignlanguage}[2]{{%
\expandafter\ifx\csname l@#1\endcsname\relax
\typeout{** WARNING: IEEEtran.bst: No hyphenation pattern has been}%
\typeout{** loaded for the language `#1'. Using the pattern for}%
\typeout{** the default language instead.}%
\else
\language=\csname l@#1\endcsname
\fi
#2}}
\providecommand{\BIBdecl}{\relax}
\BIBdecl

\bibitem{Stefan:2021:multiuserRIS}
N.~S. Perovi{\'c}, L.-N. Tran, M.~Di~Renzo, and M.~F. Flanagan, ``On the
  achievable sum-rate of the {RIS-aided} {MIMO} broadcast channel,'' in
  \emph{Proc. International Workshop on Signal Processing Advances in Wireless
  Communications (SPAWC)}.\hskip 1em plus 0.5em minus 0.4em\relax IEEE, 2021,
  pp. 571--575.

\bibitem{di2019smart}
M.~Di~Renzo \emph{et~al.}, ``Smart radio environments empowered by
  reconfigurable {AI} meta-surfaces: An idea whose time has come,''
  \emph{EURASIP J. Wireless Commun. and Netw.}, vol. 2019, no.~1, pp. 1--20,
  2019.

\bibitem{di2020smart}
------, ``Smart radio environments empowered by reconfigurable intelligent
  surfaces: How it works, state of research, and road ahead,'' \emph{{IEEE} J.
  Sel. Areas Commun.}, vol.~38, no.~11, pp. 2450--2525, Nov. 2020.

\bibitem{perovic2020achievable}
N.~S. Perovi{\'c} \emph{et~al.}, ``Achievable rate optimization for {MIMO}
  systems with reconfigurable intelligent surfaces,'' \emph{{IEEE} Trans.
  Wireless Commun.}, vol.~20, no.~6, pp. 3865--3882, Jun. 2021.

\bibitem{zhang2019capacity}
S.~Zhang and R.~Zhang, ``{Capacity characterization for intelligent reflecting
  surface aided MIMO communication},'' \emph{{IEEE} J. Sel. Areas Commun.},
  vol.~38, no.~8, pp. 1823--1838, Aug. 2020.

\bibitem{perovic2019channel}
N.~S. Perovi{\'c} \emph{et~al.}, ``Channel capacity optimization using
  reconfigurable intelligent surfaces in indoor {mmWave} environments,'' in
  \emph{Proc. IEEE Int. Conf. on Communications (ICC)}, 2020, pp. 1--7.

\bibitem{perovic2020optimization}
------, ``Optimization of {RIS}-aided {MIMO} systems via the cutoff rate,''
  \emph{{IEEE} Wireless Commun. Lett.}, vol.~10, no.~8, pp. 1692--1696, Aug.
  2021.

\bibitem{nguyen2021spectral}
N.~T. Nguyen \emph{et~al.}, ``Spectral efficiency optimization for hybrid
  relay-reflecting intelligent surface,'' in \emph{Proc. IEEE Int. Conf. on
  Communications Workshops (ICC Workshops)}.\hskip 1em plus 0.5em minus
  0.4em\relax IEEE, 2021, pp. 1--6.

\bibitem{wu2019intelligent}
Q.~Wu and R.~Zhang, ``Intelligent reflecting surface enhanced wireless network
  via joint active and passive beamforming,'' \emph{{IEEE} Trans. Wireless
  Commun.}, vol.~18, no.~11, pp. 5394--5409, Nov. 2019.

\bibitem{yu2020robust}
X.~Yu \emph{et~al.}, ``Robust and secure wireless communications via
  intelligent reflecting surfaces,'' \emph{{IEEE} J. Sel. Areas Commun.},
  vol.~38, no.~11, pp. 2637--2652, Nov. 2020.

\bibitem{kammoun2020asymptotic}
Q.-U.-A. Nadeem \emph{et~al.}, ``Asymptotic max-min {SINR} analysis of
  reconfigurable intelligent surface assisted {MISO} systems,'' \emph{{IEEE}
  Trans. Wireless Commun.}, vol.~19, no.~12, pp. 7748--7764, Dec. 2020.

\bibitem{huang2020reconfigurable}
C.~Huang \emph{et~al.}, ``{Reconfigurable intelligent surface assisted
  multiuser MISO systems exploiting deep reinforcement learning},''
  \emph{{IEEE} J. Sel. Areas Commun.}, vol.~38, no.~8, pp. 1839--1850, Aug.
  2020.

\bibitem{zhi2021two}
K.~Zhi \emph{et~al.}, ``Two-timescale design for reconfigurable intelligent
  surface-aided massive {MIMO} systems with imperfect {CSI},'' \emph{IEEE
  Trans. Inf. Theory}, 2022, {Early access}.

\bibitem{van2021reconfigurable}
T.~Van~Chien \emph{et~al.}, ``Reconfigurable intelligent surface-assisted
  cell-free massive {MIMO} systems over spatially-correlated channels,''
  \emph{{IEEE} Trans. Wireless Commun.}, vol.~21, no.~7, pp. 5106--5128, Jul.
  2022.

\bibitem{ni2021resource}
W.~Ni \emph{et~al.}, ``{Resource allocation for multi-cell IRS-aided NOMA
  networks},'' \emph{{IEEE} Trans. Wireless Commun.}, vol.~20, no.~7, pp.
  4253--4268, Jul. 2021.

\bibitem{pan2020multicell}
C.~Pan \emph{et~al.}, ``Multicell {MIMO} communications relying on intelligent
  reflecting surfaces,'' \emph{{IEEE} Trans. Wireless Commun.}, vol.~19, no.~8,
  pp. 5218--5233, Aug. 2020.

\bibitem{zhang2021joint}
Z.~Zhang and L.~Dai, ``A joint precoding framework for wideband reconfigurable
  intelligent surface-aided cell-free network,'' \emph{{IEEE} Trans. Signal
  Process.}, vol.~69, pp. 4085--4101, Jun. 2021.

\bibitem{he2020multiple}
C.~He \emph{et~al.}, ``Multiple intelligent reflecting surfaces assisted
  interference coordination in multi-cell {MU-MIMO} communications,''
  \emph{arXiv preprint arXiv:2009.13899}, 2020.

\bibitem{xu2021sum}
K.~Xu \emph{et~al.}, ``On the sum-rate of {RIS}-assisted {MIMO} multiple-access
  channels over spatially correlated rician fading,'' \emph{{IEEE} Trans.
  Commun.}, vol.~69, no.~12, pp. 8228--8241, Dec. 2021.

\bibitem{you2021reconfigurable}
L.~You \emph{et~al.}, ``Reconfigurable intelligent surfaces-assisted multiuser
  {MIMO} uplink transmission with partial {CSI},'' \emph{{IEEE} Trans. Wireless
  Commun.}, vol.~20, no.~9, pp. 5613--5627, Sep. 2021.

\bibitem{abrardo2021intelligent}
A.~Abrardo \emph{et~al.}, ``Intelligent reflecting surfaces: Sum-rate
  optimization based on statistical position information,'' \emph{{IEEE} Trans.
  Commun.}, vol.~69, no.~10, pp. 7121--7136, Oct. 2021.

\bibitem{ning2021terahertz}
B.~Ning \emph{et~al.}, ``{Terahertz multi-user massive MIMO with intelligent
  reflecting surface: Beam training and hybrid beamforming},'' \emph{{IEEE}
  Trans. Veh. Technol.}, vol.~70, no.~2, pp. 1376--1393, Feb. 2021.

\bibitem{Caire:ZFDPC:2003}
G.~Caire and S.~Shamai, ``On the achievable throughput of a multiantenna
  {Gaussian} broadcast channel,'' \emph{{IEEE} Trans. Inf. Theory}, vol.~49,
  no.~7, pp. 1691--1706, Jul. 2003.

\bibitem{yu2001trellis}
W.~Yu and J.~M. Cioffi, ``Trellis precoding for the broadcast channel,'' in
  \emph{Proc. GLOBECOM'01. IEEE Global Telecommunications Conference},
  vol.~2.\hskip 1em plus 0.5em minus 0.4em\relax IEEE, 2001, pp. 1344--1348.

\bibitem{Vishwanath:duality_achievable:2003}
S.~Vishwanath \emph{et~al.}, ``Duality, achievable rates and sum-rate capacity
  of {Gaussian MIMO} broadcast channels,'' \emph{{IEEE} Trans. Inf. Theory},
  vol.~49, no.~10, pp. 2658--2668, Oct. 2003.

\bibitem{Jindal:IterativeWF:BC:2005}
N.~Jindal \emph{et~al.}, ``Sum power iterative water-filling for multi-antenna
  {Gaussian} broadcast channels,'' \emph{{IEEE} Trans. Inf. Theory}, vol.~51,
  no.~4, pp. 1570--1580, Apr. 2005.

\bibitem{Nam:beamdesign:ZFDPC:2012}
L.-N. Tran \emph{et~al.}, ``Beamformer designs for {MISO} broadcast channels
  with zero-forcing dirty paper coding,'' \emph{{IEEE} Trans. Wireless
  Commun.}, vol.~12, no.~3, pp. 1173--1185, Mar. 2013.

\bibitem{zhang2020intelligent}
S.~Zhang and R.~Zhang, ``Intelligent reflecting surface aided multi-user
  communication: Capacity region and deployment strategy,'' \emph{{IEEE} Trans.
  Wireless Commun.}, vol.~69, no.~9, pp. 5790--5806, Sep. 2021.

\bibitem{Weingarten:CapacityRegion:MU_MIMO:2006}
H.~Weingarten \emph{et~al.}, ``The capacity region of the {Gaussian}
  multiple-input multiple-output broadcast channel,'' \emph{{IEEE} Trans. Inf.
  Theory}, vol.~52, no.~9, pp. 3936--3964, Sep. 2006.

\bibitem{Yu:SumCapacity:MIMO_BC:Decomposition:2006}
W.~Yu, ``Sum-capacity computation for the {Gaussian} vector broadcast channel
  via dual decomposition,'' \emph{{IEEE} Trans. Inf. Theory}, vol.~52, no.~2,
  pp. 754 --759, Feb. 2006.

\bibitem{Dhillon2011}
I.~S. Dhillon \emph{et~al.}, ``{Nearest neighbor based greedy coordinate
  descent},'' in \emph{Proc. Advances in Neural Information Processing Systems
  24 (NIPS 2011)}, 2011, pp. 1--9.

\bibitem{beck2009fast}
A.~Beck and M.~Teboulle, ``A fast iterative shrinkage-thresholding algorithm
  for linear inverse problems,'' \emph{SIAM journal on imaging sciences},
  vol.~2, no.~1, pp. 183--202, 2009.

\bibitem{yu2019miso}
X.~Yu \emph{et~al.}, ``{MISO} wireless communication systems via intelligent
  reflecting surfaces,'' in \emph{Proc. IEEE/CIC International Conference on
  Communications in China (ICCC)}.\hskip 1em plus 0.5em minus 0.4em\relax IEEE,
  2019, pp. 735--740.

\bibitem{Condat2016}
L.~Condat, ``{Fast projection onto the simplex and the $\pmb \ell_{1}$ ball},''
  \emph{Math. Program.}, vol. 158, no. 1-2, pp. 575--585, Jul. 2016.

\bibitem{tang2020wireless}
W.~Tang \emph{et~al.}, ``Path loss modeling and measurements for reconfigurable
  intelligent surfaces in the millimeter-wave frequency band,'' \emph{{IEEE}
  Trans. Commun.}, vol.~70, no.~9, pp. 6259--6276, Sep. 2022.

\bibitem{luo2008dynamic}
Z.-Q. Luo and S.~Zhang, ``Dynamic spectrum management: Complexity and
  duality,'' \emph{{IEEE} J. Sel. Areas Commun.}, vol.~2, no.~1, pp. 57--73,
  Feb. 2008.

\bibitem{Nam:GreedyScheduling:SZFDPC:2010}
L.-N. Tran and E.-K. Hong, ``Multiuser diversity for successive zero-forcing
  dirty paper coding: Greedy scheduling algorithms and asymptotic performance
  analysis,'' \emph{{IEEE} Trans. Signal Process.}, vol.~58, no.~6, pp.
  3411--3416, Jun. 2010.

\bibitem{pan2020intelligent}
C.~Pan \emph{et~al.}, ``Intelligent reflecting surface aided {MIMO}
  broadcasting for simultaneous wireless information and power transfer,''
  \emph{{IEEE} J. Sel. Areas Commun.}, vol.~38, no.~8, pp. 1719--1734, Aug.
  2020.

\bibitem{Hong2017}
M.~Hong \emph{et~al.}, ``{Iteration complexity analysis of block coordinate
  descent methods},'' \emph{Math. Program.}, vol. 163, no.~1, pp. 85--114, May
  2017.

\bibitem{bolte2014proximal}
J.~Bolte \emph{et~al.}, ``Proximal alternating linearized minimization for
  nonconvex and nonsmooth problems,'' \emph{Math. Program.}, vol. 146, no.~1,
  pp. 459--494, 2014.

\end{thebibliography}

\end{document}